\documentclass[useAMS,usenatbib]{mn2e}

\usepackage{graphicx}
\usepackage{rotating}
\usepackage{color}
\usepackage{url}
\usepackage{pifont}

\title[Spectroscopic Study of NGC~6811]{Spectroscopic Study of the Open Cluster NGC~6811 
\thanks{
Based on observations made with the Nordic Optical Telescope,
operated on the island of La Palma jointly by Denmark, Finland,
Iceland, Norway, and Sweden, in the Spanish Observatorio del
Roque de los Muchachos of the Instituto de Astrofis\'ica de
Canarias.
}}
\author[J.~Molenda-\.Zakowicz et~al.]{J.~Molenda-\.Zakowicz$^{1}$\thanks{E-mail: molenda@astro.uni.wroc.pl}, K.~Brogaard$^{2,3}$, E.~Niemczura$^{1}$, 
M.~Bergemann$^{4}$, \and A.~Frasca$^{5}$, T.~Arentoft$^{2}$, and F.~Grundahl$^{2}$\\
$^{1}$ Astronomical Institute, University of Wroc{\l}aw, ul.\,Kopernika 11, 51-622 Wroc{\l}aw, Poland\\
$^{2}$ Stellar Astrophysics Centre (SAC), Dept. of Physics and Astronomy, Aarhus University, Ny Munkegade 1520, 8000 Aarhus C, \\
Denmark \\
$^{3}$ Department of Physics and Astronomy, University of Victoria, P.O. Box 3055, Victoria, BC V8W 3P6, Canada \\
$^{4}$ Max-Planck-Institute for Astrophysics, Karl-Schwarzschild-Str. 1, D-85741 Garching, Germany \\
$^{5}$ INAF - Catania Astrophysical Observatory, Via S. Sofia 78, 95123 Catania, Italy } 

\begin{document}

\date{Accepted 1988 December 15. Received 1988 December 14; in original form 1988 October 11}

\pagerange{\pageref{firstpage}--\pageref{lastpage}} \pubyear{2002}

\maketitle

\label{firstpage}

\begin{abstract}
The NASA space telescope {\it Kepler} has provided unprecedented time-series observations which have revolutionised the field of asteroseismology, i.e. the use
of stellar oscillations to probe the interior of stars. The {\it Kepler}-data include observations of stars in open clusters, which are particularly interesting 
for asteroseismology. One of the clusters observed with {\it Kepler} is NGC~6811, which is the target of the present paper. However, apart from high-precision 
time-series observations, sounding the interiors of stars in open clusters by means of asteroseismology also requires accurate and precise atmospheric parameters 
as well as cluster membership indicators for the individual stars. We use medium-resolution ($R \sim 25,000$) spectroscopic observations, and three independent 
analysis methods, to derive effective temperatures, surface gravities, metallicities, projected rotational velocities and radial velocities, for 15~stars in the 
field of the open cluster NGC~6811. We discover two double-lined and three single-lined spectroscopic binaries. Eight stars are classified as either certain or 
very probable cluster members, and three stars are classified as non-members. For four stars, cluster membership could not been assessed. Five of the observed 
stars are G-type giants which are located in the colour-magnitude diagram in the region of the red clump of the cluster. Two of these stars are surely 
identified as red clump stars for the first time. For those five stars, we provide chemical abundances of 31 elements. The mean radial-velocity of NGC~6811 is 
found to be $+6.68 \pm 0.08$~km\,s$^{-1}$ and the mean metallicity and overall abundance pattern are shown to be very close to solar with an 
exception of Ba which we find to be overabundant.
\end{abstract}


\section{Introduction}
\label{intro}

NGC~6811 ($\alpha_{2000} = 19^h 37^m 10^s,$ $\delta_{\rm 2000} = +46^o 22^{\prime} 30^{\prime \prime}$) is one of four open clusters in the field of view of 
the NASA space telescope {\it Kepler} \citep{borucki2003, koch2010}. According to the WEBDA\footnote{WEBDA is a web site at 
{http://www.univie.ac.at/webda} dedicated to stellar clusters in the Galaxy and the Magellanic Clouds. In this paper we use the numbering system of that database.} 
data base, NGC~6811 is an intermediate-age ($\log (age) = 8.799$), slightly reddened ($E(B-V)=0.160$~mag) open cluster located at a distance of 1215~pc. It is 
known to host several pulsating variable stars of the $\gamma$~Dor and $\delta$~Sct type \citep{vancauteren2005, luo2009, debosscher2011, uytterhoeven2011}, stars 
showing solar-like oscillations \citep{hekker2011, stello2011, corsaro2012}, as well as other types of variable stars.

The first spectroscopic studies of NGC~6811 date back to \citet{becker1947} and \citet{lindoff1972} who used objective prisms to measure spectral types of 45 
stars. Subsequent spectroscopic observations by \citet{mermilliod1990, glushkova1999, mermilliod2008, frinchaboy2008} yielded radial-velocities ($RV$) of selected 
stars which allowed them to derive the mean $RV$ of the cluster ($\sim7$~km\,s$^{-1}$) and to verify cluster membership for stars identified as candidates by 
\citet{sanders1971, dias2002, kharchenko2004} on the basis of proper motions and the location in colour--magnitude diagrams. Those studies, however, did not 
provide the mean metallicity of the cluster which has been measured only recently by means of three different methods: spectroscopy \citep{wong2009}, an analysis 
of the $\log T_{\rm eff}$-$\log L$ diagram \citep{hekker2011}, and photometry \citep{janes2013}. All these authors find NGC~6811 to have a sub-solar metallicity, 
however, the range of the reported values ($-0.7 < \rm [Fe/H] < -0.1$~dex) is too large for computing accurate evolutionary and asteroseismic models of the 
cluster members.

The satellite {\it Kepler} has provided high-precision time-series photometry in the {\it Kepler} $Kp$ band of stars in the field of NGC~6811 for the time-period 
between March~2009 and May~2013, in either short (1-min) or long (30-min) cadence mode.  Although the main focus of the {\it Kepler} mission is planet hunting, the 
{\it Kepler} data can be used for asteroseismic studies as well. In asteroseismology, observed stellar oscillations are combined with stellar models and used to 
probe the internal structure and estimate properties of stars. For this, stars in clusters are of particular interest, because such stars are thought to have 
formed from the same cloud of interstellar gas and dust, and are expected to have similar chemical compositions, space velocities, distances and ages, thus 
limiting the number of free parameters in modelling their structure and evolution. In NGC~6811, asteroseismic studies have been carried out on red giant stars 
\citep{stello2011, corsaro2012}, based on {\it Kepler} photometry.

\begin{table*}
\caption{The individual radial velocity measurements of the programme stars.}
\label{rv-individual-6811}
\begin{tabular}{p{1.0cm}p{1.1cm}p{1.2cm}p{1.4cm}p{0.7cm}p{0.7cm}p{0.7cm}p{1.2cm}p{0.6cm}p{0.4cm}p{1.7cm}p{0.6cm}p{1.0cm}}
\hline\hline\noalign{\smallskip}
WEBDA           & KIC      &$\alpha _{\rm 2000}$ &$\delta _{\rm 2000}$ & $Kp$ &$J$  &$K_{\rm s}$& HJD      & $\rm T_{\rm exp}$ & S/N & $RV\pm\sigma$  & mem? & rem.  \\
                &          & [h:m:s]          & [deg:m:s]         &[mag] &[mag]& [mag]     & +2450000 & [s]               &     & [km\,s$^{-1}]$   \\  
\hline
\hspace{1pt} 24 & 9655101   & 19:36:57.13 &+46:22:42.6 & 10.983 &~~9.541 &~~8.994 & 4280.4872 & 1800 &~~70 & ~~~~7.49$\pm$0.53 & yes? & SB1?\\ \vspace{4pt} 
		&           &             &            &        &        &        & 4282.4906 & 1800 &~~70 & ~~~~7.36$\pm$0.51 &      & SL-like      \\
\hspace{1pt} 32 & 9655167   & 19:37:02.68 &+46:23:13.1 & 11.063 &~~9.599 &~~9.019 & 4280.5114 & 1800 &~~70 & ~~~~5.20$\pm$0.41 & yes? & SB1\\ \vspace{4pt} 
		&           &             &            &        &        &        & 4282.5148 & 1800 &~~70 & ~~~~5.37$\pm$0.49 &      & SL-like \\
\hspace{1pt} 33 & 9716220   & 19:37:05.46 &+46:24:58.5 & 11.880 & 11.243 & 11.099 & 4689.5689 & 1800 &~~60 & ~~~~3.56$\pm$0.98 & yes  & $\delta$\,Sct\\ \vspace{4pt} 
		&           &             &            &        &        &        & 4690.4987 & 1800 &~~60 & ~~~~4.62$\pm$1.07 &       &     \\
\hspace{1pt} 54 & 9655438   & 19:37:25.23 &+46:19:35.7 & 12.259 & 11.469 & 11.301 & 4689.4524 & 1800 &~~50 & ~~~~5.52$\pm$2.08 & yes  & $\gamma$\,Dor     \\ \vspace{4pt} 
		&           &             &            &        &        &        & 4690.4514 & 1800 &~~50 & ~~~~3.14$\pm$2.29 &      & hybrid    \\
\hspace{1pt} 68 & 9655187   & 19:37:04.21 &+46:18:07.7 & 11.517 & 10.967 & 10.843 & 4280.5390 & 1800 &~~50 & ~~~~7.74$\pm$0.32 &    ? & SB2\\ \vspace{4pt} 
	        &           &             &            &        &        &        & 4282.5389 & 1800 &~~50 & ~~~~7.06$\pm$0.20 &      & E      \\ 
  113           & 9655514   & 19:37:32.10 &+46:19:15.0 & 11.495 & 10.784 & 10.645 & 4280.7125 & 1800 &~~50 & ~~~~0.10$\pm$0.75 &    ? & $\delta$\,Sct    \\ \vspace{4pt}
		&           &             &            &        &        &        & 4282.5631 & 1800 &~~50 & ~~~~0.51$\pm$0.47        &     \\
  133           & 9716090   & 19:36:55.81 &+46:27:37.7 & 11.133 &~~9.671 &~~9.086 & 4280.5736 & 1800 &~~70 & ~~~~6.79$\pm$0.36 &  yes & SL-like   \\ \vspace{4pt} 
		&           &             &            &        &        &        & 4282.5875 & 1800 &~~70 & ~~~~6.80$\pm$0.43        &     \\
  173$^{\star}$ & 9594100\,A& 19:36:55.98 &+46:15:18.5 & 13.014 & 12.145 & 11.897 & 4689.4758 & 1500 &~~30 & ~~~~8.52$\pm$0.55 &   no & SB2\\
		& 9594100\,B&             &            &        &        &        & 4689.4758 & 1500 &~~30 & $-$52.13$\pm$1.06 &      & $\gamma$\,Dor   \\
		& 9594100\,A&             &            &        &        &        & 4690.4751 & 1800 &~~30 & ~~~~7.79$\pm$0.32 &      &    \\ \vspace{4pt}
		& 9594100\,B&             &            &        &        &        & 4690.4751 & 1800 &~~30 & $-$57.73$\pm$1.51 &      &    \\
  218           & 9716667   & 19:37:48.08 &+46:27:25.3 & 12.634 & 11.532 & 11.392 & 4689.4266 & 1800 &~~50 & ~~~~6.56$\pm$1.46 &    ? & SB1\\ \vspace{4pt} 
                &           &             &            &        &        &        & 4690.4270 & 1800 &~~50 & ~~~~7.08$\pm$1.51 &      & $\delta$\,Sct      \\             
  408           & 9715189   & 19:35:31.42 &+46:27:45.0 &~~9.993 &~~7.118 &~~6.013 & 4280.4644 & 1500 & 100 & $-$24.94$\pm$1.20 &   no &     \\ \vspace{4pt}
		&           &             &            &        &        &        & 4282.4434 & 1500 & 100 & $-$25.06$\pm$1.12        &     \\
  471           & 9776739   & 19:37:22.09 &+46:32:50.6 & 10.904 &~~9.489 &~~8.927 & 4280.5980 & 1800 &~~80 & ~~~~6.88$\pm$0.46 &  yes & SL-like \\ \vspace{4pt} 
		&           &             &            &        &        &        & 4282.6283 & 1800 &~~80 & ~~~~6.84$\pm$0.39        &     \\
  483           & 9532903   & 19:37:50.18 &+46:07:46.5 & 10.936 &~~9.464 &~~8.901 & 4280.6387 & 1800 &~~70 & ~~~~6.39$\pm$0.46 &  yes & SL-like \\ \vspace{4pt} 
		&           &             &            &        &        &        & 4282.6526 & 1800 &~~70 & ~~~~6.39$\pm$0.42        &     \\
  489           & 9594857   & 19:37:58.76 &+46:14:19.4 & 11.021 & 10.199 & 10.017 & 4280.6633 & 1800 &~~70 &~~$-$2.06$\pm$0.44 &    ? & SB1?\\ \vspace{4pt} 
	        &           &             &            &        &        &        & 4282.6768 & 1800 &~~70 & ~~~~5.74$\pm$0.60 &      & $\delta$\,Sct            \\           
  528           & 9777532   & 19:38:31.06 &+46:31:34.1 & 10.940 & 10.255 & 10.102 & 4280.6874 & 1800 &~~70 & ~~~~6.37$\pm$0.24 &  yes & rotation/    \\ \vspace{4pt} 
		&           &             &            &        &        &        & 4282.7011 & 1800 &~~70 & ~~~~6.78$\pm$0.21 &      & activity    \\
  K1            & 9895798   & 19:35:38.55 &+46:42:25.8 &~~8.900 &~~6.130 &~~5.033 & 4280.4475 &~~600 &~~90 & $-$23.39$\pm$0.24 &   no &     \\ 
		&           &             &            &        &        &        & 4282.4278 &~~600 & 100 & $-$23.42$\pm$0.23        &     \\
\hline               
\multicolumn{12}{l}{$^{\rm \star}$ The WEBDA number, the equatorial coordinates, and the $Kp, J$, and $K_{\rm s}$ magnitudes refer to both components.}\\
\end{tabular}
\end{table*}

Apart from the high-precision time-series photometry from {\it Kepler}, asteroseismic modelling requires precise values of effective temperature ($T_{\rm eff}$), 
surface gravity ($\log g$), and iron abundance ($\rm [Fe/H]$). The latter is often used as a proxy for the metallicity of the star, although preferably
the whole abundance pattern should be known. Those values are provided in the 
\textit{Kepler Input Catalog} (KIC) \citep{brown2011} only for a fraction of the stars, and those values are not precise enough for asteroseismic modelling. A 
re-determination from ground-based spectroscopic or photometric observations is therefore required \citep[see][]{molenda2010, brown2011}.

Extensive programmes for ground-based follow-up observations of \textit{Kepler} asteroseismic targets, aiming at deriving their atmospheric parameters have been 
developed \citep[see][]{Uytterhoeven2010a, Uytterhoeven2010b}. In this paper, we report results from an analysis of observations acquired before the launch of 
\textit{Kepler}. The goal of the observations was to derive $T_{\rm eff}$, $\log g$, chemical abundances, radial velocities ($RV$), projected rotational 
velocities ($v\sin i$), and membership status in the cluster for 15 stars. We also aimed at obtaining the mean radial-velocity and metallicity of the cluster.

The paper is organised as follows. In Section~\ref{targets} we outline the method of selecting our targets, and the observations and data reductions are described 
in Section~\ref{observations}. Radial velocities, cluster membership information and the mean $RV$ of NGC~6811 are provided in Section~\ref{rv-sec}. In 
Section~\ref{section:atmos}, we derive and present atmospheric parameters for our targets, and in Section~\ref{ab-analysis} we derive abundances of 31 elements 
for the five red clump stars in our sample. The global cluster parameters are estimated and discussed in Section~\ref{cluster_param} and, finally, a summary is 
provided in  Section~\ref{sum}.

\section{Target selection}
\label{targets}

We set out to observe a sample of stars which were likely cluster members, and for which we could obtain spectra with sufficiently high signal-to-noise to 
perform our analysis aiming at determining physical parameters for the individual stars and the global parameters for NGC~6811. We, therefore, selected 15~bright 
stars in the field of NGC~6811, of which most are classified as cluster members by \citet{dias2002, sanders1971, kharchenko2004, mermilliod1990}, and which have 
spectral types ranging from early F to late K. Two stars expected to be cool giants, stars 408 and K1 (see below), were included in our sample in order to 
determine cluster membership. In such a sample, different types of pulsating stars would be expected. Therefore, these stars were also submitted in the first run 
of the {\it Kepler} Asteroseismic Science Consortium (KASC)\footnote{\url{http://astro.phys.au.dk/KASC}} proposals in September~2008 as candidates for 
\textit{Kepler} asteroseismic targets (proposal P01\_27) and were accepted for observations. All but one were observed by \textit{Kepler} since the beginning of 
the mission, i.e. since the so-called quarter Q0, which started on~2 May~2009, through quarter Q16 which ended on~8 April~2013. KIC~9895798 was observed only in 
quarters Q0 and Q1, ending on 15~June~2009. All targets were observed in the short cadence mode for at least one quarter\footnote{\url 
{http://keplergo.arc.nasa.gov/ArchiveSchedule.shtml}}.

The data acquired in Q1 allowed \citet{debosscher2011} or \citet{uytterhoeven2011} to detect photometric variability in star~33 and~218 which were classified as 
$\delta$~Sct-type pulsators, star~54 which was classified as $\gamma$~Dor by \citet{debosscher2011} and as hybrid pulsator by \citet{uytterhoeven2011}, and 
star~173 which was classified as $\gamma$~Dor. Our sample also includes stars~113 and~489, discovered by \citet{vancauteren2005} to be $\delta$~Sct-type variables, 
and star~68 discovered by \citet{Watson2006} to be an eclipsing binary (E). Five G-type giants from our sample, i.e. stars~24, 32, 133, 471, and~483, show 
solar-like (hereafter 'SL-like') oscillations in {\it Kepler} photometry \citep[see][]{hekker2011, stello2011, corsaro2012}. Finally, star~528 shows variability due to 
rotation or activity \citep{uytterhoeven2011}.

Our programme stars are listed in Table~\ref{rv-individual-6811}. In columns~1 through~5, we give their KIC and WEBDA numbers, their equatorial coordinates, and 
the \textit{Kepler} $Kp$ magnitudes. The last column provides information about photometric variability of the star. For KIC~9895798, which does not have 
a~WEBDA number, we use a short name 'K1'.

\section{Observations and Data Reduction}
\label{observations}

The spectroscopic observations were carried out at Observatorio del Roque de los Muchachos, La Palma, Spain, on four nights:~28 and~30 June~2007, and~10 and~11 
August~2008. For each star, we acquired two spectra separated by one or two nights. We used the 2.56-m Nordic Optical Telescope (NOT) equipped with the Fibre-fed 
Echelle Spectrograph FIES and the NIMO back illuminated $2048 \times2048$ CCD\,42-40. The spectrograph has a resolving power of $R=25,000$ and covers the entire 
spectral range of 370-730~nm without gaps in a single, fixed setting. The readout noise was 3.3~$\bar e$ and a gain of 0.71~$\bar e$/ADU.

For the basic reductions, we used the NOAO/IRAF package\footnote{IRAF is distributed by the National Optical Astronomy Observatory, which is operated by the 
Association of Universities for Research in Astronomy, Inc.}. This reduction included subtracting a bias frame, trimming the images, and correcting for flat 
field and scattered light. The spectra were extracted with the IRAF \textsc{apall} task, and wavelength-calibrated with respect to the spectra of a Th-Ar 
comparison lamp which were taken before and after each exposure on the programme stars.

In columns 8--10 of Table~\ref{rv-individual-6811}, we list the Heliocentric Julian Day (HJD) of the observations, the exposure time, and the signal-to-noise 
ratio of each observation.

\section{Radial velocities}
\label{rv-sec}

The radial velocities were derived using the \textsc{IRAF fxcor} package.  When deriving the $RV$-values, we used spectra of two different objects for reference. 
For the late-type stars~24, 32, 133, 471, and 483, we used the $RV$-standard star 31\,Aql \citep[G8\,IV, $RV = -100.35$~km\,s$^{-1}$,][]{udry1999} which was
observed on the same nights as the programme stars. For the remaining early-type stars, we used a synthetic spectrum for $T_{\rm eff} = 7500$~K, $\log g = 
4.0$~dex, and solar metallicity, computed with the line-blanketed LTE codes \textsc{ATLAS9} and \textsc{SYNTHE} \citep{kurucz93}.

We measured the $RV$ in each order of the echelle spectrum and then computed a weighted mean, averaging the measurements from all orders. For each order $i$, we 
adopted a statistical weight $w_i = \sigma _i^{-2}$ where the values of $\sigma _i$ were computed with the \textsc{fxcor} task by taking into account the 
height of the fitted peak and the antisymmetric noise \citep[see][]{tonry79}. The uncertainty of the weighted mean was computed as described by \citet{topping72}. 
The radial velocities of our programme stars and their standard deviations are given in column~11 of Table~\ref{rv-individual-6811}.

\begin{figure}
\includegraphics[width=8.5cm,angle=0]{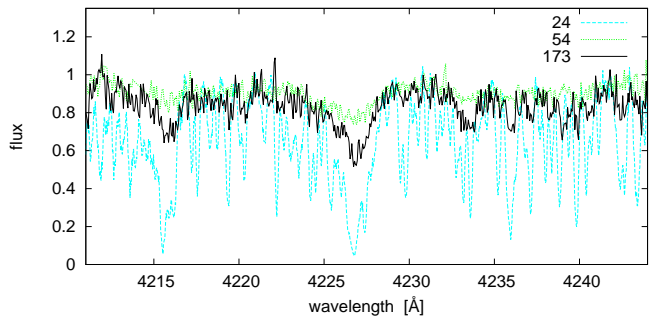}
\caption[]{A comparison of a part of the spectrum of the SB2 star~173 (thick black line), the hot star~54 (dotted green line), and the cool star~24 (dashed blue 
line).}
\label{star173}
\end{figure}

\subsection{Stars with variable radial velocity}
\label{rv-var}

Our observations of each target star were separated by only one or two nights, which is not optimal for detecting $RV$ variability. Indeed, we only measure 
$RV$ changes for two stars:~173 and~489. Star~173, which we discover to be a double-lined spectroscopic binary (SB2), shows spectral features of a hot star 
superimposed on the spectrum of a cool star. This is shown in Fig.~\ref{star173}, where we plot a part of the spectra for stars~173, 24 (a cool star), and 54 
(a hot star). The $RV$ values of the~A and~B components of star~173 measured on 10 and~11 August~2008 are similar, which is consistent with the $~$1-day period 
shown by this star in the \textit{Kepler} photometry \citep{uytterhoeven2011}.

For star~489, we discovered significant changes in $RV$. Furthermore, both our measured values, $-2.06 \pm 0.44$ and $+5.74 \pm 0.60$~km\,s$^{-1}$, differ from 
the $RV$ value reported by \citet{frinchaboy2008} of $+20.01 \pm 2.54$~km\,s$^{-1}$, which suggests that star~489 is a spectroscopic binary. Indeed, its spectral 
lines show a slight asymmetry, but that may be caused either by another star, making star~489 an SB2 system, or by star spots or other surface variations such as 
pulsations. The latter possibility is very probable because the star is a $\delta$~Sct-type variable \citep{vancauteren2005}. Therefore, we caution the reader 
not to put to much trust in our spectroscopic results derived for star~489 under the assumption of a single star. A similar caution should be adopted also for the 
other hot, fast-rotating stars in our sample (see Sect.~\ref{results:F} and Table~\ref{table:atmos:F}).

\begin{figure}
\includegraphics[width=8.5cm,angle=0]{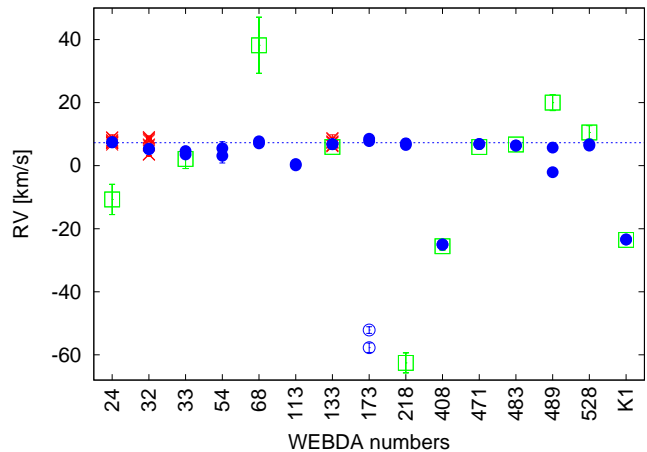}
\caption[]{Blue dots indicate $RV$ values of single stars, SB1 systems, and star~173~A from Table~\ref{rv-individual-6811}. Blue open circles indicate the $RV$ 
values of star~173~B; green squares are the $RV$ values of stars observed by \citet{frinchaboy2008}, and red crosses are $RV$ values of stars observed by 
\citet{mermilliod2008}. The dashed line indicates the mean $RV$ of NGC~6811, $+6.68$~km\,s$^{-1}$, as derived in this paper.}
\label{rv-ngc6811}
\end{figure}

For the four stars described below, the spectroscopic binarity was either already known, or has been discovered by comparing our $RV$ measurements with those 
available in the literature (see Fig.~\ref{rv-ngc6811}).

Star~24 has $RV$-values which are constant within the $1\sigma$ error-bars in our observations and they are consistent with the nine $RV$ values reported by 
\citet{mermilliod2008}. However, they differ from the single $RV$ measurement obtained by \citet{frinchaboy2008}, $-10.71 \pm 4.80$~km\,s$^{-1}$. The observations 
by \citet{mermilliod2008} cover a time-span of 18 years. During that period, the radial velocity of star~24 has never fallen outside the range from $+6.68\pm0.45$ 
to $+8.97\pm0.80$~km\,s$^{-1}$. Therefore, the measurement reported by \citet{frinchaboy2008} is unexpected. We classify star~24 as a suspected single-lined 
spectroscopic binary (SB1?).

Star~32 has been discovered as a spectroscopic binary by \citet{mermilliod1990}. The $RV$ values derived here, $+5.20 \pm 0.41$ and $+5.37 \pm 
0.49$~km\,s$^{-1}$, fall in the range of the $RV$-values reported by \citet{mermilliod2008}, which was from $+3.49 \pm 0.53$ to $+9.07 \pm 0.38$~km\,s$^{-1}$.

Star~68 has been discovered as an eclipsing binary by \citet{Watson2006} and is listed in the \textit{Kepler} eclipsing binary catalogue by \citet{prsa2011}. 
The \textit{Kepler} light curve of this star reveals a contact system with partial eclipses of similar, but not identical, depths. Because of the near equal depth 
of the eclipses, one would expect star~68 as an SB2 system. However, our spectra are separated in time by about half an orbital period, which amounts to 
4.41599~days \citep{prsa2011}, and show only one component with a high rotational velocity. We expect that this is due to an unfortunate timing of our 
observations, which happens to coincide with the times of eclipses and the corresponding $RV$ crossing of the binary components. Further observations could 
confirm this. The $RV$ value reported for this target by \citet{frinchaboy2008}, $+38.18 \pm 8.89$~km\,s$^{-1}$, is significantly higher than the $RV$ values 
measured here. However, the fact that our observations most likely include the combined spectra of two fast rotating components, one of which is partially 
eclipsed, makes a spectroscopic analysis very difficult.

Star~218 does not show $RV$ variability in our observations ($RV = +6.56 \pm 1.46$ and $+7.08 \pm 1.51$~km\,s$^{-1}$), but the $RV$ value reported by 
\citet{frinchaboy2008} of $-62.54 \pm 3.17$~km\,s$^{-1}$ is significantly different from our measurements. This star has been classified as a $\delta$-Sct-type 
variable by \citet{debosscher2011}, but an amplitude equal to almost 70~km\,s$^{-1}$ is too high to be caused only by pulsations. We classify star~218 as a 
single-lined spectroscopic binary.

The stars classified as spectroscopic binaries in this paper are indicated as such in the last column of Table~\ref{rv-individual-6811}.

\subsection{Cluster membership}
\label{membership}

The large amount of spectroscopic binaries in our sample makes it difficult, from our data, to determine if the stars are cluster members. Stars which are not 
known or suspected spectroscopic binaries, and which have $RV$ values in the range from~6 to~9~km\,s$^{-1}$ to within $3\sigma$, are classified as certain cluster 
members. This range is based on the determinations of the mean radial velocity of the cluster, as obtained by other authors. These range from $+6.03 \pm 
0.30$~km\,s$^{-1}$ by \cite{frinchaboy2008} to $+8.7$~km\,s$^{-1}$ by \cite{wong2009}.

This approach allowed us to classify stars~33, 54, 133, 471, 483, and 528 as cluster members. Stars~24 and 32 are classified as very probable cluster members. 
Star~32 is a known, and star~24 a suspected, spectroscopic binary. However, both stars have been classified as certain cluster members based on their proper 
motions, their radial velocities, and their photometric and seismic properties by other authors \citep[see][]{sanders1971, mermilliod1990, dias2002, stello2011}.

Stars~173, 408 and~K1 are classified in the present paper as non-members because their radial-velocities are significantly different from the cluster mean value. 
In the case of the SB2 star~173, we made use of the fact that the $RV$ of the A~component is very close to the mean $RV$ of the cluster, derived in the previous 
studies, while the $RV$ of the B~component is at around $-52$~km\,s$^{-1}$. This leads to the conclusion that the systemic velocity must be different from the 
mean velocity of NGC~6811, thus the stellar system does not belong to cluster.

For the SB2 star~68 and the three $\delta$~Sct-type variables, stars~113, 218 and 489, membership have not been assessed. All these stars require more observations 
in order to determine their mean radial velocity and draw conclusions on their membership to the cluster.

\begin{figure}
\includegraphics[width=8.5cm,angle=0]{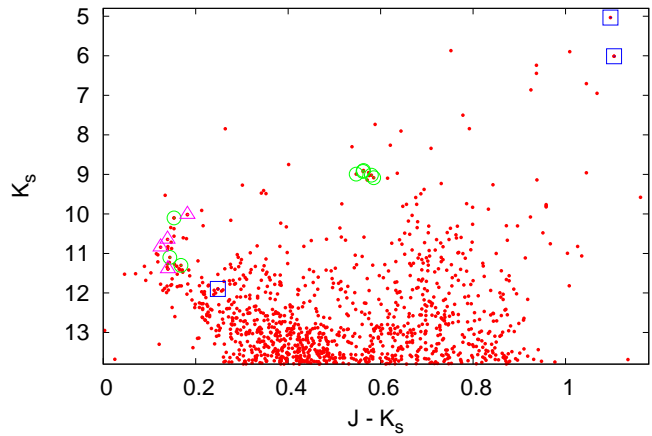}
\caption{Colour-magnitude diagram based on stars within a radius of $r=20$~arcmin from the centre of the open cluster NGC~6811. The symbols indicate, for our 
targets, certain and the very probable members of the cluster (circles), non-members (squares), and the stars for which membership is uncertain (triangles). The 
remaining stars are plotted with dots.}
\label{cmd}
\end{figure}

The membership status of our target stars to NGC~6811 is given in Table~\ref{rv-individual-6811}. The six stars classified as members 
are indicated with 'yes', the two very probable members with 'yes?', the three non-members with 'no', and the four stars without membership classification with a 
question mark. In Fig.~\ref{cmd}, we show the colour-magnitude diagram of the cluster with the positions of our targets indicated. The colour-magnitude diagram 
is based on~$J$ and~$K_S$ magnitudes adopted from the 2MASS catalogue \citep{Cutri2003}. The~$J$ and~$K_S$ magnitudes of our programme stars are reproduced in 
columns~6 and~7 in Table~\ref{rv-individual-6811}.

\subsection{Mean radial-velocity of the cluster}
\label{meanrv}

The first determination of the mean radial-velocity of NGC~6811 was provided by \citet{mermilliod1990} who used the CORAVEL spectrovelocimeters to acquire from 
three to six spectrograms of stars~24, 32, 73, 79, 101, 133, 210, 223, 234, and 237. Those stars were classified as probable cluster members by
\citet{sanders1971} who analysed proper motions of 296 stars in the field of NGC~6811 and concluded that 97 of the stars most probably belong to the cluster. 
\citet{mermilliod1990} confirmed the cluster membership for stars~24, 32, 101, and 133, and derived the mean $RV$ of the cluster, $+7.14\pm 0.26$~km\,s$^{-1}$, 
using stars~24, 101, and 133. Star~32 was not used because those authors discovered it to be a spectroscopic binary system. Eighteen years later, 
\citet{mermilliod2008} re-determined the mean $RV$ of NGC~6811 obtaining the value $+7.28 \pm 0.19$~km\,s$^{-1}$ from the same three stars and a more extended 
set of observations. 

A slightly lower value of $+6.05 \pm 0.95$~km\,s$^{-1}$ was obtained by \citet{kharchenko2005} who used literature $RV$ values of two
cluster members. That value is close to $+6.03 \pm 0.30$~km\,s$^{-1}$ derived by \citet{frinchaboy2008} from single spectrograms of stars~33, 106, 133, 471, 483, 
TYC~3556-02634-1, and TYC~3556-00370-1 acquired with the WIYN~3.5-m telescope at the Kitt Peak National Observatory.

The most recent determination of the mean radial-velocity of NGC~6811, $+8.7$~km\,s$^{-1}$, has been reported by \citet{wong2009} who observed 1157 stars in the 
field of the cluster with the Hectochelle spectrograph at the Multiple Mirror Telescope and classified~139 of them as candidate cluster members.

In this paper, we derive the radial velocity of NGC~6811 to be $+6.68 \pm 0.08$~km\,s$^{-1}$, by computing a weighted mean of the $RV$s of stars~133, 471, 483, 
and 528 which are neither known nor suspected spectroscopic binaries, and which are classified in this paper as cluster members. In the calculations, we did not 
include stars~33 and~54, because these are fast rotators causing less precise $RV$ values and, for the hot stars, our evaluation of whether the spectra contain 
spectral lines from a second component is less certain.

\section{Atmospheric Parameters}
\label{section:atmos}

\subsection{Methods}
\label{methods}

We use three independent methods of spectroscopic analysis for deriving atmospheric parameters and projected rotational velocities of our target stars. These are 
implemented in the codes \textsc{FITSUN, SME,} and \textsc{ROTFIT} which adopts different approaches for deriving $T_{\rm eff}$, $\log g$, [Fe/H], and $v\sin i$. 
Each of these methods as suffers from different limitations, as described below. The purpose of choosing these three significantly different methods was to 
benefit from diversification of approaches in order to evaluate the overall uncertainties in the results as outlined by, e.g., Smalley (in prep.).

\subsubsection{ROTFIT}
\label{methods:rotfit}

The \textsc{ROTFIT} code has been developed by \citet{Frasca2003, Frasca2006} to perform an automatic classification of the spectral type and the luminosity 
class, and to derive the projected rotational velocity of late-type stars. Subsequently, the method was developed to allow simultaneous determination of 
$T_{\rm eff}$, $\log g$, $\rm [Fe/H]$, $v\sin i$, and the MK type of those stars. 

One of the advantages of \textsc{ROTFIT} is that, unlike other methods of spectroscopic analysis, it can be applied also to spectra of low resolution, low 
signal-to-noise ratio, and $v\sin i$ exceeding 20~km~s$^{-1}$. The properties and limitations of \textsc{ROTFIT} have been described and discussed in detail by 
\citet{molenda2013}. Briefly, the method consists in comparing spectra of the programme stars with a library of spectra of the reference stars. The latter, 
consisting of 221~stars listed by \citet{molenda2013}, was constructed from the high-resolution ($R = 42,000$), high signal-to-noise spectra of slowly rotating 
stars from the ELODIE archive \citep{prugniel2001} for which the atmospheric parameters are measured with relatively high accuracy.

The uncertainties of $T_{\rm eff}$, $\log g$, $\rm[Fe/H]$, and $v \sin i$ are the standard errors of the respective weighted means to which the average 
uncertainties of the stellar parameters of the reference stars, i.e. $\sigma_{T_{\rm eff}}= 88$~K, $\sigma_{\log g}=0.21$\,dex, and $\sigma_{\rm [Fe/H]}=
0.21$\,dex, are added in quadrature \citep[see][]{molenda2013}. 

\subsubsection{FITSUN}
\label{methods:atlas}

In this method, spectral synthesis based on the least squares optimisation algorithm are used to determine the atmospheric parameters, the projected rotational 
velocity and radial velocity of stars. To perform the spectrum analysis, a pre-computed grid of atmospheric models is necessary. For the stars analysed in this 
paper, we derived the line-blanketed LTE models by using the code \textsc{ATLAS9} by \citet{kurucz93}. The grid covered the range in $T_{\rm eff}$ from 4750 to 
6000~K with a step of 50~K and from 6000 to 9000~K with a step of 100~K. The range in $\log g$ is from 2.70 to 4.50~dex with a step of 0.10~dex. The grid 
has been computed for solar metallicity with the abundances of elements from \citet{grevesse98}. The spectrum synthesis code \textsc{FITSUN} uses the 
\textsc{SYNTHE} suite of programs by \citet{kurucz93} which allow to compute synthetic spectra. Both \textsc{ATLAS9} and \textsc{SYNTHE} have been ported under 
\textsc{GNU Linux} by \citet{sbordone05} and are available online\footnote{\url{http://atmos.obspm.fr/}}.

The detailed description of the method and the performance of the code \textsc{FITSUN} has been provided by \citet{niemczura09}. This method allows for deriving 
atmospheric parameters of stars by carrying out spectral synthesis for spectral features chosen according to the spectral type of the target. The line list by
Kurucz\footnote{\url{http://kurucz.harvard.edu/linelists.html}} is used. We start by examining the observed spectrum and adopting an initial 
set of atmospheric parameters of the targets, either by visual inspection of selected lines in their spectra (F-type stars) or by adopting the values derived by 
means of other methods (i.e. the code \textsc{ROTFIT}). Then, we build a dense grid of atmosphere models, large enough to cover the expected range of the 
atmospheric parameters for our targets. Before running the code \textsc{FITSUN}, the observed spectra must be prepared for the analysis. This includes 
normalisation to the level of continuum, deciding which spectral features will be considered for analysis and dividing the spectrum into parts. The length of 
those parts depends mainly on the values of $v\sin i $ of the star. Then, \textsc{FITSUN} makes use of \textsc{SYNTHE} and the adopted atmosphere model to 
calculate the synthetic spectrum in the selected parts, and uses the least-squares method to calculate the abundances of elements, the radial velocity, and the 
projected rotational velocity of the targets, as described by \citet{niemczura09}. If necessary, \textsc{FITSUN} corrects on-the-fly the continuum placement of 
the selected parts of spectra. The eventual usefulness of this method depends on the projected rotational velocity of the star and the correct continuum 
placement. 

\subsubsection{SME}
\label{methods:SME}

The package Spectroscopy Made Easy\footnote{\url{http://www.stsci.edu/~valenti/sme.html}} (\textsc{SME}) by \citet{valenti1996} is an \textsc{IDL} program which 
allows to fit the observed spectrum with a synthetic one and to determine the atmospheric parameters of the programme star. In \textsc{SME}, the spectrum 
synthesis and the $\chi ^2$-minimisation in the spectral regions defined by the input masks are carried out on-the-fly. In these computations, we used the 
MARCS\footnote{\url{http://marcs.astro.uu.se/}} model atmospheres by \citet{gustafsson2008}. When defining the masks, we selected only those iron lines for which 
the complete NLTE grids of abundance corrections are available. In total, we used about~60 diagnostic Fe~I and Fe~II transitions. The line list was built from the 
SIU line list which is a compilation by different groups \citep{korn2003, grupp2004a, grupp2004b, bergemann2008, bergemann2010, bergemann2012a, bergemann2012b, 
onehag2011, shi2014} and applied in different spectroscopic studies \citep[see, e.g.][]{bergemann2011}. The stellar parameters were determined iteratively, 
exploring the full parameter-space in $T_{\rm eff}$, $\log g$, [Fe/H], the micro-turbulence velocity ($\xi_{\rm t}$) and the macro-turbulence velocity 
($v_{\rm mac}$). The influence of NLTE effects was treated using the model atom and the grids presented by \citet{bergemann2012c, lind2012}. 

The uncertainties of the stellar parameters produced by \textsc{SME} are combined estimates stemming from several sources. First, \textsc{SME} accounts for the 
signal-to-noise of a spectrum while iteratively searching for the most probable solution in the full parameter space. The robustness of the final values of
the parameters is assessed by perturbing the initial guesses and reiterating until convergence. These are the internal uncertainties of the method. Second, we 
include systematic errors which have been determined from the analysis of a reference high-resolution stellar sample \citep[see][Bergemann et~al., 
in prep.]{bergemann2012c} for which the parameters obtained by independent methods such as interferometry and asteroseismology are available. The mean difference 
between those values and the results produced by \textsc{SME} are adopted for a measure of the systematic error for a given spectral type. The individual 
uncertainties are combined in quadrature and their maximum is adopted as the final uncertainty of the parameter under consideration. 

\subsection{Results}
\label{results}

The atmospheric parameters obtained by means of the three methods described in Section~\ref{methods} are discussed below, separately for the F-, G-, and M-type 
stars. For each star and each method, the values which we report are means calculated from the results obtained from two individual spectra.

\subsubsection{G-type stars}
\label{results:G}

\begin{table}
\begin{center}
\caption{Seismic parameters of G-type stars.}
\label{table:seismic}
\begin{tabular}{rrrcrrr}
\hline\hline\noalign{\smallskip}
WEBDA (KIC)    & $\Delta \nu \pm \sigma$ & $\nu _{\rm max} \pm \sigma $  & $\log g_{\rm seis.}$ \\ 
               & [$\mu$Hz]    & [$\mu$Hz] & [dex]\\ \hline
  24 (9655101) & 7.86$\pm$0.04     &  98.2$\pm$2.4 &  2.90\\
  32 (9655167) & 8.07$\pm$0.04     & 100.3$\pm$8.7 &  2.91\\
 133 (9716090) & 8.56$\pm$0.06     & 101.4$\pm$5.9 &  2.92\\
 471 (9776739) & 7.93$\pm$0.16     &  93.4$\pm$9.0 &  2.88\\
 483 (9532903) & 7.69$\pm$0.16     &  96.3$\pm$4.5 &  2.90\\
\hline
\end{tabular}
\end{center}
\end{table}

The G-type stars~24, 32, 133, 471, and 483 observed by us in the field of NGC~6811 form a small, distinct group located at $K_S \simeq 9$~mag, $(J-K_S) \simeq 
0.6$~mag in the colour-magnitude diagram shown in Fig.~\ref{cmd}. Stars~24, 32, and 133, have been classified as red clump (RC) by \citet{mermilliod1990} who 
inferred that from the colour-magnitude diagram of the cluster. This classification was then confirmed by \citet{corsaro2012} who analysed the asteroseismic 
properties of these stars. Our analysis shows that also stars~471 and~483 should be classified as RC. They are cluster members (see Sect.~\ref{membership}) and 
they fall in the same region of the $K_S - (J-K_S)$ diagram as stars~24, 32, and 133 (see Fig.~\ref{cmd}). Moreover, they have very similar values of the 
atmospheric parameters as the three other RC stars (see Table~\ref{table:atmos:giants}). Unfortunately, neither star~471 nor~483 was included in the sample 
analysed by \citet{corsaro2012}.

However, all five stars show solar-like oscillations which was discovered in the \textit{Kepler} data by \citet{hekker2011} and \citet{stello2011}. As showed 
in Table~\ref{table:seismic}, the asteroseismic parameters of these stars are very similar which further indicates that they are all RC-stars. We have computed 
asteroseismic values of $\log g$ for these five stars, in order to check the accuracy of our spectroscopic values $\log g$ values. We made use of the fact that 
the asteroseismic values of $\log g$ are not only precise but also accurate, at least at metallicities close to and higher than solar, even if the mass derived 
from the asteroseismic scaling relations seems to be slightly overestimated \citep[see][]{brogaard2012, sandquist2013, frandsen2013}. 
We used the asteroseismic parameters derived for stars~24, 32 and~133 by \citet{corsaro2012}, and for stars~471 and~483 by \citet{hekker2011}, kindly provided to 
us by the first authors of those two papers. We reproduce these parameters in Table~\ref{table:seismic} where the symbols $\Delta \nu$ and $\nu _{\rm max}$ stand for
the large separation between consecutive overtones and the frequency of maximum power of the oscillations, respectively \citep[see, e.g.,][]{stello2009}. We note
that the differences between uncertainties of those values reported by \citet{hekker2011} and \citet{corsaro2012} result from the different lengths of the
time-series analysed by these two groups (nearly a year by \citet{hekker2011} and more than~19 months by \citet{corsaro2012}). In the computations, we used the 
asteroseismic scaling relations for mass and radius \citep[eqns.~3 and~4 from][]{miglio2012} and the values of $T_{\rm eff}$ derived with \textsc{FITSUN} 
(Table~\ref{table:atmos:giants}). In principle, the spectroscopic analysis should be iterated after using the derived values of $T_{\rm eff}$ to calculate new asteroseismic 
values of $\log g$. However, since changes to the values of $\log g$ were already less than 0.01~dex in the first iteration, further iterations were not needed.
The resulting values of $\log g_{\rm seism.}$ are provided in the fourth column of Table~\ref{table:seismic}. A conservative uncertainty estimate for these values 
is $\pm 0.04$~dex.

\begin{table*}
\begin{center}
\caption{Atmospheric parameters of G-type stars; the uncertainties of the derived values are given in the text.} 
\label{table:atmos:giants}
\begin{tabular}{rclrrccrrrc}
\hline\hline\noalign{\smallskip}
\multicolumn{11}{c}{\bf\textsc FITSUN}\\ \hline
WEBDA (KIC)      &$T_{\rm eff}$& $\log g$ & $\rm [Fe/H]$ & $v\sin i$ &$\xi_{\rm t}$& $T_{\rm eff}$& $\log g _{\rm seis.}$ & $\rm [Fe/H]$ & $v\sin i$ &$\xi_{\rm t}$\\
                 & [K]         & [dex]    & [dex]        & [km\,s$^{-1}$] & [km\,s$^{-1}$] & [K]         & [dex]    & [dex]        & [km\,s$^{-1}$] & [km\,s$^{-1}$] \\ \smallskip
                 & \multicolumn{5}{c}{\hrulefill \raisebox{-2pt}{$\log g$ free} \hrulefill}   
                 & \multicolumn{5}{c}{\hrulefill \raisebox{-2pt}{$\log g_{\rm seis.}$ fixed} \hrulefill}\\
     24 (9655101) & 5050 & 3.00 &   0.06 & 4.8& 1.00   & 5000 & 2.90 &   0.00 & 4.8& 1.10\\
     32 (9655167) & 5050 & 2.90 &   0.09 & 5.1& 1.10   & 5050 & 2.91 &   0.09 & 5.1& 1.10\\
    133 (9716090) & 5100 & 2.90 &   0.07 & 4.5& 1.10   & 5100 & 2.92 &   0.07 & 4.5& 1.10\\
    471 (9776739) & 4950 & 2.90 &   0.04 & 4.7& 1.10   & 4950 & 2.88 &   0.04 & 4.7& 1.10\\
    483 (9532903) & \textemdash & \textemdash & \textemdash   & \textemdash  & \textemdash  & 5050 & 2.90 &   0.04& 4.7& 1.10\\ 
\hline\noalign{\smallskip}
\multicolumn{11}{c}{\bf\textsc SME} \\ \hline
WEBDA (KIC)     &$T_{\rm eff}$& $\log g$ & $\rm [Fe/H]$ & $v\sin i$ &$\xi_{\rm t}$; $v_{\rm mac}$ & $T_{\rm eff}$& $\log g _{\rm seis.}$ & $\rm [Fe/H]$ & $v\sin i$ & $\xi_{\rm t}$; $v_{\rm mac}$\\
                & [K]         & [dex]    & [dex]        &  [km\,s$^{-1}$] &[km\,s$^{-1}$] & [K]         & [dex]    & [dex]        & [km\,s$^{-1}$] & [km\,s$^{-1}$]\\ \smallskip
                & \multicolumn{5}{c}{\hrulefill \raisebox{-2pt}{$\log g$ free} \hrulefill} 
                & \multicolumn{5}{c}{\hrulefill \raisebox{-2pt}{$\log g_{\rm seis.}$ fixed} \hrulefill}\\
    24 (9655101) & 5083 & 3.05 &   0.08 & 0.1&  1.11; 3.85 & 5005 & 2.90 &   0.03&0.1&  1.11; 3.85\\   
    32 (9655167) & 4939 & 2.95 &   0.06 & 4.1&  1.11; 3.85 & 4924 & 2.91 &   0.05&4.1&  1.11; 3.85\\  
   133 (9716090) & 5005 & 2.97 &   0.05 & 0.1&  1.11; 3.84 & 4980 & 2.92 &   0.02&0.1&  1.11; 3.84\\  
   471 (9776739) & 4922 & 2.81 &   0.03 & 0.1&  1.11; 3.87 & 4952 & 2.88 &   0.05&0.1&  1.11; 3.87\\  
   483 (9532903) & 5066 & 3.02 &   0.09 & 0.1&  1.11; 3.91 & 5008 & 2.90 &   0.05&0.1&  1.11; 3.91\\  
\hline\noalign{\smallskip}
&\multicolumn{5}{c}{\bf\textsc ROTFIT} & \multicolumn{3}{c}{\bf\textit{Kepler} Input Catalog}\\ \hline
 WEBDA (KIC)     &$T_{\rm eff}$& $\log g$& [Fe/H] & $v\sin i$ & MK &  $T_{\rm eff}$& $\log g$& $\rm [Fe/H]$\\
                 & [K]         & [dex]    & [dex] & [km\,s$^{-1}$] & type  & [K]         & [dex]    & [dex]\\ \smallskip
                 & \multicolumn{5}{c}{\hrulefill \raisebox{-2pt}{$\log g$ free} \hrulefill}  & \multicolumn{3}{c}{\hrulefill \raisebox{-2pt}{\textit{griz} photometry} \hrulefill} \\
    24 (9655101) & 5135 & 3.18 &$-$0.05 & 0.1 & G8\,III& 5078 & 2.9  &   0.05 \\
    32 (9655167) & 5103 & 3.17 &$-$0.04 & 0.1 & G8\,III& 5034 & 2.9  &   0.00 \\
   133 (9716090) & 5130 & 3.20 &$-$0.04 & 0.1 & G8\,III& 5042 & 2.8  &$-$0.10 \\
   471 (9776739) & 5095 & 3.11 &$-$0.04 & 0.1 & G8\,III& 5146 & 2.9  &$-$0.09 \\
   483 (9532903) & 5120 & 3.21 &$-$0.04 & 0.1 & G8\,III& 5119 & 2.5  &   0.06 \\
\hline
\hline
\end{tabular}
\end{center}
\end{table*}

Below, we present the atmospheric parameters obtained for these five stars independently with the codes \textsc{FITSUN, SME}, and \textsc{ROTFIT}. In case of 
\textsc{FITSUN} and \textsc{SME}, apart from the purely spectroscopic analysis, we carried out additional computations with $\log g$ fixed to the asteroseismic
values from Table~\ref{table:seismic}. In Table~\ref{table:atmos:giants}, where we present the results, the parameters obtained from purely spectroscopic analyses 
are labelled '$\log g$ free' while those obtained for $\log g$ fixed to the asteroseismic value are labelled '$\log g$ fixed'. Table~\ref{table:atmos:giants}
also provides the photometric values of the atmospheric parameters of our targets reproduced from the KIC.

\begin{table*}
\begin{center}
\caption{Mean atmospheric parameters of G-type stars and the biases between the methods.}
\label{table:biases}
\begin{tabular}{lrrrrrrrr}
\hline\hline\noalign{\smallskip}
Parameter                         & \textsc{FITSUN}& \textsc{SME}& \textsc{ROTFIT}
& $\Delta_{\tiny\textsc{(FITSUN - SME)}}$ 
& $\Delta_{\tiny\textsc{(FITSUN - ROTFIT)}}$ 
& $\Delta_{\tiny\textsc{(ROTFIT - SME)}}$\\ \hline
$\rm \langle T_{\rm eff} \rangle _{\,\log g\, free}$ [K]               & $5038\pm32$   & $5003\pm32$   & $5117\pm8$      & $35\pm45$      & $-79\pm33$     & $114\pm33$\\
$\rm \langle \log g      \rangle _{\,\log g\, free}$ [dex]             & $2.95\pm0.03$ & $2.96\pm0.04$ & $3.17\pm0.02$   & $-0.01\pm0.05$ & $-0.22\pm0.04$ & $0.21\pm0.05$\\
$\rm \langle [Fe/H]      \rangle _{\,\log g\, free}$ [dex]             & $0.07\pm0.01$ & $0.06\pm0.01$ & $-0.04\pm0.002$ & $ 0.01\pm0.02$ & $0.11\pm0.01$  & $-0.10\pm0.01$\\
$\rm \langle T_{\rm eff} \rangle _{\,\log g_{\rm seis.}\, fixed}$ [K]  & $5030\pm25$   & $4974\pm16$   & \textemdash     & $56\pm30$      & \textemdash    & \textemdash\\
$\rm \langle \log g_{\rm seis.}\rangle $ [dex]                         & $2.90\pm0.01$ & $2.90\pm0.01$ & \textemdash     & \textemdash    & \textemdash    & \textemdash\\
$\rm \langle [Fe/H]      \rangle _{\,\log g_{\rm seis.}\, fixed}$ [dex]& $0.05\pm0.02$ & $0.04\pm0.01$ & \textemdash     & $0.01\pm0.02$  & \textemdash    & \textemdash\\
\hline
\end{tabular}
\end{center}
\end{table*}

\begin{itemize}
\item \textsc{FITSUN}
    
\noindent
The atmospheric parameters obtained with \textsc{FITSUN} are provided in the top part of Table~\ref{table:atmos:giants}. The left side of the Table, labelled 
'$\log g$ free', lists the values of $T_{\rm eff}$, $\log g$, $\rm [Fe/H]$, and $v\sin i$, resulting from a purely spectroscopic analysis. The values of $\rm 
[Fe/H]$  and $T_{\rm eff}$ have been derived from isolated, unblended lines of Fe~I. The right side of Table~\ref{table:atmos:giants} lists the atmospheric 
parameters and $v\sin i$ obtained with $\log g$ fixed to the asteroseismic value for which a dedicated grid of model atmospheres has been prepared. In both 
computations, the value of $v_{\rm mac}$ has been set to zero. For star~483 for which we detected only one useful line of Fe\,II we present only 
the atmospheric parameters derived for the fixed, asteroseismic value of $\log g$. The steps in our grid of model atmospheres, i.e. 50~K in $T_{\rm eff}$, 
0.1~dex in $\log g$, and 0.5~km\,s$^{-1}$ in $\xi_t$, define the uncertainties of these parameters. The typical uncertainty of the value of $\rm [Fe/H]$ is 
0.20~dex (c.f. Table~\ref{elements:abs}.) The typical uncertainty of $v\sin i$ is $\pm 1.0$~km\,s$^{-1}$.

\item \textsc{SME}

\noindent
The NLTE results obtained with the code \textsc{SME} and the grid of NLTE corrections by \citet{lind2012} are provided in the middle part of 
Table~\ref{table:atmos:giants} separately for the free and the fixed values of $\log g$. The values of $\xi _{\rm t}$ and $v_{\rm mac}$ have been obtained from 
a calibration relation by Bergemann et~al., (in prep.) The uncertainties of the computed values are 150~K in $T_{\rm eff}$, 0.15~dex in $\log g$, 0.10~dex in 
[Fe/H], and 0.5~km\,s$^{-1}$ in and $v\sin i$. The uncertainties for $\xi_{\rm t}$ and $v_{\rm mac}$ have not been provided. Both quantities are ad-hoc 
parameters introduced to approximate non-thermal broadening and do not have a physical measurable equivalent. In 3D hydrodynamic simulations, turbulent velocity 
fields are stochastic and do not take a single constant value throughout a stellar atmosphere.

\item \textsc{ROTFIT}

\noindent
The results obtained with \textsc{ROTFIT} are provided at the bottom left of Table~\ref{table:atmos:giants}. They have been obtained only for the free $\log g$ 
because \textsc{ROTFIT} does not allow to fix any of the atmospheric parameters \citep[see][]{Frasca2003, Frasca2006, molenda2013}. The standard errors of 
the obtained atmospheric parameters are 120~K in $T_{\rm eff}$, 0.22~dex in $\log g$ and [Fe/H], and 0.4~km\,s$^{-1}$ in $v \sin i$. 

\item KIC
    
\noindent
The photometric values of $T_{\rm eff}$, $\log g$, and $\rm [Fe/H]$ from the KIC catalogue have been provided at the bottom right of 
Table~\ref{table:atmos:giants}. The precision of these parameters is 200~K in $T_{\rm eff}$, and 0.5~dex in $\log g$ and $\rm [Fe/H]$ \citep{brown2011}. 
\end{itemize}

The values of $T_{\rm eff}$, $\log g$, and $\rm [Fe/H]$ derived for G-type stars by means of the three methods used in this paper agree with each other 
to within $1\sigma$ error bars. The mean values of $T_{\rm eff}$, $\log g$, and $\rm [Fe/H]$ of G-type stars in NGC~6811, calculated from the individual 
values from Table~\ref{table:atmos:giants}, are provided in Table~\ref{table:biases}. The same table provides also biases between \textsc{FITSUN}, \textsc{SME}, 
and \textsc{ROTFIT}. The uncertainties of the reported values are standard deviations of the mean.

The biases listed in Table~\ref{table:biases} show that \textsc{ROTFIT} produced a temperature scale which is higher by around 70~K comparing to \textsc{FITSUN}
and around 110~K comparing to \textsc{SME}, while \textsc{SME} produced a temperature scale which is lower by around 30~K comparing to \textsc{FITSUN}. For
the fixed values of $\log g$, the last difference amounts to around 70~K. 

The mean values of $\log g$ and $\rm [Fe/H]$ produced by \textsc{FITSUN} and \textsc{SME} agree nicely with each other. They are also in a very 
good agreement with the asteroseismic values of $\log g$. A similarly good agreement between the spectroscopic and asteroseismic values of $\log g$ has been 
reported by \citet{bruntt2012} and \citet{thygesen2012} who used the code \textsc{VWA} \citep{bruntt2010} to perform spectroscopic analysis of 93~solar-type stars
and 82~red giants, and detected only a handful of outliers. However, one cannot just expect a general good agreement between spectroscopic and asteroseismic 
$\log g$ values. As a relevant example, quite discordant values of spectroscopic $\log g$ were obtained by \citet{molenda2013} with the code 
\textsc{ARES+MOOG} \citep{santos2004, sousa2006, sousa2008, sousa2011a, sousa2011b} for stars~24, 32, and~133 which are on average 0.5~dex higher than the 
asteroseismic values reported in this paper. (We note that the values of $\log g$ obtained by \citet{molenda2013} with the code \textsc{ROTFIT} agree well with 
those derived with ROTFIT in this paper.)

Apart from a higher temperature scale, \textsc{ROTFIT} produced also values of $\log g$ which are higher by 0.20-0.25~dex, and values of $\rm [Fe/H]$ which are 
lower by 0.06-0.10~dex in comparison with the two other methods. The simplest explanation for that is a weak correlation between the fitted parameters 
$T_{\rm eff}$ and $\log g$ which has a different behaviour depending on the adopted analysis code and, most importantly, on the analysed spectral range. 
Therefore, it is not surprising that \textsc{FITSUN} and \textsc{SME} which use unblended Fe\,I and Fe\,II lines agree better with each other than with 
\textsc{ROTFIT} which analyses entire spectral segments containing also strong and broad absorption lines. The observed discrepancy may be also due to the grid 
of templates which, being composed of real star spectra, cannot cover all the regions of the parameter-space with equal density \citep[see figure~2 
in][]{molenda2013}. On the other hand, we note that the biases reported in Table~\ref{table:biases} may be lower or even reversed when \textsc{ROTFIT} is compared 
to other spectroscopic or photometric methods \citep[see figures~4 and~6 in][]{molenda2013}. Therefore, our results support the conclusion of \citet{molenda2013} 
that the present accuracy of determinations of the atmospheric parameters of solar-type stars is not better than $\pm 150$~K in $T_{\rm eff}$, $\pm 0.15$~dex in 
[Fe/H], and $\pm 0.3$~dex in $\log g$ when no asteroseismic $\log g$ values are available.

When explaining the sources of differences between the results produced by \textsc{FITSUN} and \textsc{SME}, multiple reasons can be indicated (see, e.g., 
Smalley in prep.). Below, we list the most important ones \citep[the following comparisons does not concern \textsc{ROTFIT} because the latter code uses for an 
input a grid of atmospheric parameters of reference stars that were computed by various authors and by means of different methods; see][]{molenda2013}:
\begin{itemize}
    \item The choice of model atmospheres: \textsc{FITSUN} uses the plane-parallel LTE model atmospheres by Kurucz while \textsc{SME} uses the 
        spherically-symmetric MARCS model atmospheres;
    \item The adopted atomic data: \textsc{FITSUN} uses the Kurucz atomic and molecular line list (see Sect.~\ref{methods:atlas}) while 
          \textsc{SME} uses the updated line list from SIU spectrum synthesis code (see Sect.~\ref{methods:SME});
    \item The reference to the solar abundances: \textsc{FITSUN} uses the reference by \citet{grevesse98} while \textsc{SME}, the reference by \citet{grevesse07}
        which is the source of a slight difference in solar abundances adopted in the two codes ($\log (\rm Fe) = 7.50$ in \textsc{FITSUN} and $\log 
        (\rm Fe)=7.45$ in \textsc{SME});
    \item The procedure of deriving the atmospheric parameters: \textsc{FITSUN} uses an iterative approach to derive the atmospheric parameters while 
        \textsc{SME}, which also solves for stellar parameters iteratively, exploring the full parameter space, derives the values of $T_{\rm eff}$, $\log g$, 
        and $\rm [Fe/H]$ simultaneously.
\end{itemize}
These are accompanied with other, secondary factors which are altogether sufficient to produce noticeable differences between the values of $T_{\rm eff}$, 
$\log g$, and $\rm [Fe/H]$ derived by means of different methods. As shown by, e.g., \citet{metcalfe2010, gomez2013}, or and Guzik et~al., in prep., such 
differences for stars similar to the Sun or slightly hotter can easily exceed 100~K in $T_{\rm eff}$ and 0.6~dex in $\log g$ \citep[c.f. the results
obtained for stars 24, 32, and~133 by][]{molenda2013}.

Since all the G-type stars from Table~\ref{table:atmos:giants} have been classified in this paper as cluster members, their mean metallicity $\langle \rm [Fe/H] 
\rangle$ can be adopted for the mean metallicity of the cluster. These mean values, calculated separately for each method, are provided in 
Table~\ref{table:biases}. They range from $-0.04\pm0.002$~dex obtained with \textsc{ROTFIT} to $0.06\pm0.01$~dex obtained with \textsc{SME} from the fully 
spectroscopic analysis, i.e. the '$\log g$ free' case. These values agree with what is expected for a Galactic open cluster located at the galactocentric
radius $R_{\rm GC} = 8.3$~kpc 
and an age of about 1~Gyr \citep[see Sect.~\ref{cluster_param} and figures~9 and~10 in][]{pancino2010}. Previous determinations of $\rm [Fe/H]$ of NGC~6811 
were lower. They ranged from $-0.1$~dex, derived by \citet{wong2009} from observations acquired with the Hectochelle spectrograph on the Multiple Mirror 
Telescope, through $-0.19$~dex, derived by \citet{janes2013} from photometric observations and fitting isochrones to the cluster colour-magnitude diagram, to
a value as low as $\rm -0.3 > [Fe/H] > -0.7$~dex, as proposed by \citet{hekker2011} from an analysis of the position of four cluster red giants in the $\log 
T_{\rm eff}$--$\log L$ diagram. Neither the results obtained in this paper nor those reported by \citet{molenda2013} confirm such a low metallicity of NGC~6811.
In Sect.~\ref{cluster_param}, we provide explanations for the deviating results of \citet{hekker2011} and \citet{janes2013}.

All our codes find the G-type stars to be slow rotators. \textsc{FITSUN} finds the values of $v\sin i$ of our targets to fall in the range from 3.7 to 
4.4~km\,s$^{-1}$. This range is confirmed by \textsc{SME} only for star~32; the remaining stars are found by that code to have $v\sin i = 0.1$~km\,s$^{-1}$.
The difference in $v\sin i$ as derived by \textsc{FITSUN} and \textsc{SME} is caused by the different assumptions on $v_{\rm mac}$. Also \textsc{ROTFIT} yields 
$v\sin i = 0.1$~km\,s$^{-1}$ for all the five stars but in that case, the values of $v\sin i$ may be underestimated; when computing $v\sin i$, 
\textsc{ROTFIT} assumes that the reference stars do not rotate while in reality they do. However, at the resolution of our spectra (R=25,000) \textsc{ROTFIT} 
can neither easily resolve the effect of $v_{\rm mac}$ and $v\sin i$ on the line profiles nor measure any broadening effect smaller than around~5~km\,s$^{-1}$. 
Therefore, we conclude that the values of $v\sin i$ obtained with the three methods are consistent with each other.

\subsubsection{F-type stars}
\label{results:F}

In the case of fast rotating F-type stars, to derive $T_{\rm eff}$ and $\log g$, we used the Balmer lines, and we required the same abundances from all available Fe~I 
and Fe~II lines. Since we had low signal-to-noise spectra of fast rotating stars at our disposal, we did not detect the weak, unblended lines of iron which is 
why the determination of accurate values of $\xi_{\rm t}$ and $\log g$ was impossible. Therefore, we decided to fix the value of $\xi_{\rm t}$ to~3 
or~4~km\,s$^{-1}$, depending on the resulting [Fe/H] which we tried to keep as close as possible to the mean value of [Fe/H] of G-type stars derived with 
\textsc{FITSUN}, derive only the values of $T_{\rm eff}$, and estimate the values of $\log g$. For each of the stars the last parameter was falling in the range 
from~3.5 to~4.0~dex with the uncertainty of 0.4~dex, and with little influence on the eventual values of $T_{\rm eff}$. 

The results of this analysis are presented in Table~\ref{table:atmos:F}. The uncertainty of the values of $T_{\rm eff}$ is 100~K. The uncertainties of the values 
of $v\sin i$ are provided in the Table. Table~\ref{table:atmos:F} lists also the values of $T_{\rm eff}$ from the KIC which are available for three of our targets
and which agree with our determinations to within $1\sigma$~error bars. 

We do not provide the results of the analyses carried out with \textsc{SME} and \textsc{ROTFIT} because for the former code, the spectra were of too low quality 
while for the latter one, the stars fall too close to the limits in $T_{\rm eff}$ of the grid of the reference stars \citep[see][]{molenda2013}.

\begin{table}
\begin{center}
\caption{Atmospheric parameters of F-type stars.} 
\label{table:atmos:F}
\begin{tabular}{rrlr}
\hline\hline\noalign{\smallskip}
&\multicolumn{2}{c}{\bf FITSUN}\\ \hline
WEBDA (KIC)  &$T_{\rm eff}$& $v\sin i$     & $T_{\rm eff, KIC}$ \\
             & [K]         & [km\,s$^{-1}$]& [K]\\
\hline 
  33 (9716220) & 7600 & 172 $\pm$                        20  & 7725\\ 
  54 (9655438) & 7100 & 222 $\pm$                        16  & 6974\\ 
 113 (9655514) & 7400 & 101 $\pm$\hspace{4pt}             9  & 7441\\ 
 218 (9716667) & 7600 & 167 $\pm$                        24  & \textemdash    \\ 
 489 (9594857) & 7000 & \hspace{4pt}94 $\pm$\hspace{4pt}  9  & \textemdash    \\ 
 528 (9777532) & 7300 & \hspace{4pt}65 $\pm$\hspace{4pt}  6  & \textemdash    \\ 
\hline
\end{tabular}
\end{center}
\end{table}

\begin{table}
\begin{center}
\caption{Atmospheric parameters of M-type stars.} 
\label{table:atmos:M}
\begin{tabular}{rllrrcc}
\hline\hline\noalign{\smallskip}
&\multicolumn{4}{c}{\bf ROTFIT}\\ \hline
WEBDA (KIC)              &$T_{\rm eff}$& $\log g$& [Fe/H] & $v\sin i$ &SpT \\ 
                         & [K]         & [dex]   & [dex]  & [km\,s$^{-1}$]\\
\hline 
408 (9715189)            & 3891 & 1.75 &$-$0.03 & 0.1 & M1\,III \\ 
 K1 (9895798)            & 3868 & 1.71 &$-$0.06 & 6.1 & M1\,III \\ 
\hline\noalign{\smallskip}
&\multicolumn{4}{c}{\bf\textit{Kepler} Input Catalog}\\ \hline
WEBDA (KIC) &$T_{\rm eff}$& $\log g$& [Fe/H] \\ 
            & [K]         & [dex]   & [dex]\\
\hline
K1 (9895798)& 3356 & 0.5 & 0.58\\
\hline
\end{tabular}
\end{center}
\end{table}

\begin{figure}
\includegraphics[width=8.5cm,angle=0]{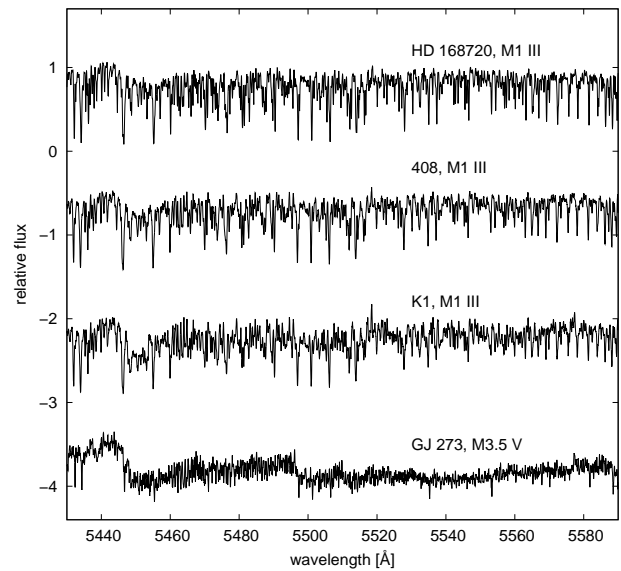}
\caption[]{Part of spectra of two \textsc{ROTFIT} reference stars HD~168720 (M1\,III) and GJ~273 (M3.5\,V), and the programme stars~408, and K1.}
\label{mstars}
\end{figure}

\subsubsection{M-type stars}
\label{results:M}

The M-type stars~408 and~K1, which are not cluster members, were analysed with the code \textsc{ROTFIT}. The values of $T_{\rm eff}$, $\log g$, $\rm [Fe/H]$, 
$v\sin i$, and the MK type obtained for them are provided in Table~\ref{table:atmos:M}. The uncertainties of the derived atmospheric parameters are 100~K in 
$T_{\rm eff}$, 0.22~dex in $\log g$ and [Fe/H], and 0.8~km\,s$^{-1}$ in $v \sin i$. However, since the two stars fall close to the edge of the grid of 
parameters of the reference stars used in \textsc{ROTFIT}, \citep[c.f. figure~3 in][]{molenda2013}, our determinations may be less accurate than the formal errors.

The KIC catalogue gives the photometric values of $T_{\rm eff}$, $\log g$, and [Fe/H] only for star~K1 which is found there to be significantly cooler, more 
evolved, and metal-rich (see Table~\ref{table:atmos:M}.) Our analysis does not confirm these values, particularly the value of $T_{\rm eff}$ (we do not discuss 
the other parameters because our grid of reference spectra contains only a few cool, low-gravity, metal-rich stars.) In order to illustrate the difference 
between the spectrum of a star of $T_{\rm eff} = 3900$ and a star of $T_{\rm eff} = 3400$~K, we plot part of the spectra of stars~408 and~K1, and the spectra of
two \textsc{ROTFIT} reference stars HD\,168720 \citep[$T_{\rm eff} = 3810$~K, M1\,III,][]{valdes2004} and GJ~273 \citep[$T_{\rm eff} = 3420$~K,
M3.5\,V,][]{cesetti2013} in Fig.~\ref{mstars}. Even a visual comparison of these spectra allows to conclude that star~K1 can not be as cool as 3350~K which proves
that the value of $T_{\rm eff}$ in the KIC is erroneous. 

We do not provide the atmospheric parameters derived for star~408 and~K1 with \textsc{FITSUN} and \textsc{SME}, because the spectral lines of both of these were
severely blended with molecular bands and are therefore not suitable for these two codes.

\section{Abundance analysis}

\label{ab-analysis}

\begin{table*}
\caption{Chemical abundances of the G-type red clump stars in NGC~6811.} 
\label{elements:abs}
\begin{tabular}{p{0.6cm}p{1.3cm}p{0.6cm}p{1.3cm}p{0.6cm}p{1.3cm}p{0.6cm}p{1.3cm}p{0.6cm}p{1.3cm}p{0.6cm}p{1.3cm}}
    \hline\hline\noalign{\smallskip}
    El.       & \multicolumn{10}{c}{WEBDA (KIC)}                                                                                        & Sun\\\noalign{\smallskip}
              & 24              &       & 32            &       & 133           &       & 471           &       & 483           &       & \\
              &    (9655101)    &       &    (9655167)  &       &     (9716090) &       &     (9776739) &       &     (9532903) &       & \\ \hline
    $_3 \rm Li$ & $1.34       $ &   [1] & $1.46       $ &   [1] & $1.37       $ &   [1] & $1.12       $ &   [1] & $1.27       $ &   [1] & $1.10\pm0.10$\\ 
 $_{11} \rm Na$ & $6.45\pm0.16$ &   [5] & $6.43\pm0.06$ &   [3] & $6.51\pm0.25$ &   [8] & $6.33\pm0.06$ &   [5] & $6.46\pm0.14$ &   [5] & $6.33\pm0.03$\\ 
 $_{12} \rm Mg$ & $7.60\pm0.22$ &  [10] & $7.61\pm0.12$ &   [3] & $7.69\pm0.11$ &   [4] & $7.32\pm0.11$ &   [8] & $7.59\pm0.29$ &   [4] & $7.58\pm0.05$\\ 
 $_{13} \rm Al$ & $6.12       $ &   [1] & $6.49       $ &   [1] & $6.18       $ &   [1] & $6.63       $ &   [1] & \textemdash   &       & $6.43\pm0.07$\\ 
 $_{14} \rm Si$ & $7.41\pm0.36$ &  [49] & $7.50\pm0.34$ &  [40] & $7.23\pm0.29$ &  [40] & $7.23\pm0.36$ &  [45] & $7.33\pm0.26$ &  [43] & $7.55\pm0.05$\\ 
 $_{16} \rm S $ & $7.33       $ &   [2] & $7.55       $ &   [2] & \textemdash   &       & \textemdash   &       & \textemdash   &       & $7.33\pm0.11$\\ 
 $_{20} \rm Ca$ & $6.38\pm0.27$ &  [35] & $6.48\pm0.24$ &  [19] & $6.50\pm0.21$ &  [21] & $6.36\pm0.22$ &  [23] & $6.23\pm0.26$ &  [23] & $6.36\pm0.02$\\ 
 $_{21} \rm Sc$ & $3.06\pm0.33$ &  [28] & $3.07\pm0.26$ &  [21] & $3.08\pm0.15$ &  [14] & $3.00\pm0.22$ &  [17] & $3.00\pm0.14$ &  [19] & $3.17\pm0.10$\\ 
 $_{22} \rm Ti$ & $4.93\pm0.27$ & [142] & $4.91\pm0.26$ & [111] & $4.95\pm0.27$ & [115] & $4.85\pm0.18$ & [124] & $4.88\pm0.18$ & [102] & $5.02\pm0.06$\\ 
 $_{23} \rm V $ & $3.85\pm0.28$ &  [67] & $3.92\pm0.18$ &  [47] & $3.93\pm0.17$ &  [55] & $3.80\pm0.22$ &  [63] & $3.85\pm0.20$ &  [51] & $4.00\pm0.02$\\ 
 $_{24} \rm Cr$ & $5.67\pm0.19$ & [110] & $5.71\pm0.19$ &  [79] & $5.67\pm0.17$ &  [87] & $5.65\pm0.23$ & [102] & $5.65\pm0.22$ & [100] & $5.67\pm0.03$\\ 
 $_{25} \rm Mn$ & $5.30\pm0.31$ &  [43] & $5.38\pm0.23$ &  [32] & $5.34\pm0.28$ &  [41] & $5.35\pm0.23$ &  [41] & $5.35\pm0.23$ &  [38] & $5.39\pm0.03$\\ 
 $_{26} \rm Fe$ & $7.51\pm0.17$ & [369] & $7.54\pm0.16$ & [191] & $7.58\pm0.14$ & [258] & $7.51\pm0.15$ & [298] & $7.51\pm0.14$ & [316] & $7.50\pm0.05$\\ 
 $_{26} \rm Fe\,I$  & $7.56\pm0.22$ &  [95] & $7.59\pm0.19$ &  [94] & $7.57\pm0.22$ & [117] & $7.54\pm0.23$ &  [89] & $7.54\pm0.19$ &  [88] &          \\ 
 $_{26} \rm Fe\,II$ & $7.56\pm0.17$ &   [6] & $7.60       $ &   [2] & $7.57\pm0.10$ &   [3] & $7.53\pm0.29$ &   [6] & $7.51       $ &   [1] &          \\ 
 $_{27} \rm Co$ & $4.86\pm0.21$ &  [65] & $4.85\pm0.19$ &  [41] & $4.88\pm0.22$ &  [44] & $4.79\pm0.25$ &  [58] & $4.82\pm0.25$ &  [62] & $4.92\pm0.04$\\ 
 $_{28} \rm Ni$ & $6.22\pm0.22$ & [115] & $6.27\pm0.19$ &  [96] & $6.22\pm0.23$ & [112] & $6.25\pm0.22$ & [111] & $6.20\pm0.20$ & [108] & $6.25\pm0.04$\\ 
 $_{29} \rm Cu$ & $4.35\pm0.22$ &   [3] & $4.42       $ &   [2] & $4.38       $ &   [2] & $4.34\pm0.36$ &   [3] & $4.37\pm0.56$ &   [3] & $4.21\pm0.04$\\ 
 $_{30} \rm Zn$ & $4.81       $ &   [1] & \textemdash   &       & $4.37       $ &   [1] & $4.72       $ &   [1] & \textemdash   &       & $4.56\pm0.08$\\ 
 $_{38} \rm Sr$ & $3.67       $ &   [1] & \textemdash   &       & $3.50\pm0.13$ &   [3] & $3.38\pm0.32$ &   [3] & $3.58       $ &   [2] & $2.97\pm0.05$\\ 
 $_{39} \rm Y $ & $2.45\pm0.18$ &  [13] & $2.36\pm0.15$ &   [8] & $2.33\pm0.17$ &   [8] & $2.32\pm0.25$ &  [18] & $2.31\pm0.19$ &  [15] & $2.24\pm0.03$\\ 
 $_{40} \rm Zr$ & $2.71\pm0.24$ &  [13] & $2.24\pm0.23$ &  [10] & $2.65\pm0.37$ &   [7] & $2.25\pm0.34$ &  [13] & $2.41\pm0.26$ &   [6] & $2.60\pm0.02$\\ 
 $_{41} \rm Nb$ & $1.66       $ &   [1] & \textemdash   &       & $1.65       $ &   [1] & \textemdash   &       & \textemdash   &       & $1.38\pm0.06$\\ 
 $_{42} \rm Mo$ & $2.11       $ &   [2] & $2.21       $ &   [2] & $2.18       $ &   [1] & $2.08       $ &   [1] & $2.13       $ &   [1] & $1.92\pm0.05$\\ 
 $_{44} \rm Ru$ & \textemdash   &       & $1.71       $ &   [1] & \textemdash   &       & $1.73       $ &   [2] & $1.82       $ &   [1] & $1.84\pm0.07$\\ 
 $_{56} \rm Ba$ & $2.90\pm0.22$ &   [5] & $2.94       $ &   [2] & $2.89       $ &   [2] & $2.92\pm0.04$ &   [3] & $2.84       $ &   [2] & $2.13\pm0.05$\\ 
 $_{57} \rm La$ & $1.50\pm0.20$ &  [15] & $1.42\pm0.07$ &   [8] & $1.56\pm0.19$ &   [6] & $1.43\pm0.29$ &  [11] & $1.36\pm0.15$ &   [8] & $1.17\pm0.07$\\ 
 $_{58} \rm Ce$ & $1.89\pm0.32$ &  [20] & $2.22\pm0.34$ &  [10] & $1.78\pm0.33$ &  [10] & $2.12\pm0.25$ &  [18] & $1.88\pm0.20$ &  [12] & $1.58\pm0.09$\\ 
 $_{59} \rm Pr$ & $1.04\pm0.47$ &   [8] & $1.03\pm0.11$ &   [4] & \textemdash   &       & $0.81\pm0.36$ &   [8] & $0.52\pm0.18$ &   [5] & $0.71\pm0.08$\\ 
 $_{60} \rm Nd$ & $1.65\pm0.30$ &  [39] & $1.63\pm0.24$ &  [20] & $1.64\pm0.22$ &  [29] & $1.69\pm0.21$ &  [39] & $1.60\pm0.22$ &  [27] & $1.50\pm0.06$\\ 
 $_{62} \rm Sm$ & $1.88\pm0.63$ &   [5] & \textemdash   &       & $1.81\pm0.32$ &   [3] & $1.68\pm0.21$ &   [7] & $1.49       $ &   [2] & $1.01\pm0.06$\\ 
 $_{64} \rm Gd$ & $1.15       $ &   [2] & \textemdash   &       & \textemdash   &       & \textemdash   &       & \textemdash   &       & $1.12\pm0.04$\\ 
 $_{66} \rm Dy$ & $1.08       $ &   [1] & \textemdash   &       & $1.25       $ &   [2] & \textemdash   &       & \textemdash   &       & $1.10\pm0.08$\\ 
\hline                                                                                                
\end{tabular}                                                                                         
\end{table*}

\begin{table*}
\begin{center}
\caption{The ratios of the abundances of elements to iron for the G-type red clump stars in NGC~6811.} 
\label{elements:rel}
\begin{tabular}{lllllll}
    \hline\hline\noalign{\smallskip}
    [X/Fe]     & \multicolumn{5}{c}{WEBDA (KIC)}                                                                  & \\\noalign{\smallskip}
               & 24 (9655101)                & 32 (9655167)                & 133 (9716090)                & 471 (9776739)               & 483 (9532903)               & cluster mean\\ \hline
 $ \rm [Li/Fe]$&\hspace{6.5pt}$ 0.18       $ &\hspace{6.5pt}$ 0.27       $ &\hspace{6.5pt}$ 0.20        $ &              $-0.02       $ &\hspace{6.5pt}$ 0.13       $ &\hspace{6.5pt}$ 0.15\pm 0.10$ \\ 
 $ \rm [Na/Fe]$&\hspace{6.5pt}$ 0.06\pm0.28$ &\hspace{6.5pt}$ 0.01\pm0.21$ &\hspace{6.5pt}$ 0.11\pm0.34 $ &              $-0.04\pm0.24$ &\hspace{6.5pt}$ 0.09\pm0.24$ &\hspace{6.5pt}$ 0.04\pm 0.11$ \\ 
 $ \rm [Mg/Fe]$&              $-0.04\pm0.32$ &              $-0.06\pm0.24$ &\hspace{6.5pt}$ 0.04\pm0.26 $ &              $-0.30\pm0.26$ &              $-0.03\pm0.35$ &              $-0.09\pm 0.12$ \\ 
 $ \rm [Al/Fe]$&              $-0.37       $ &              $-0.03       $ &              $-0.32        $ &\hspace{6.5pt}$ 0.16       $ &\hspace{6.5pt}\textemdash    &              $-0.14\pm 0.22$ \\ 
 $ \rm [Si/Fe]$&              $-0.20\pm0.43$ &              $-0.14\pm0.40$ &              $-0.39\pm0.37 $ &              $-0.36\pm0.43$ &              $-0.26\pm0.33$ &              $-0.27\pm 0.17$ \\ 
 $ \rm [S/Fe] $&              $-0.06       $ &\hspace{6.5pt}$ 0.13       $ &\hspace{6.5pt}\textemdash     &\hspace{6.5pt}\textemdash    &\hspace{6.5pt}\textemdash    &\hspace{6.5pt}$ 0.03\pm 0.09$ \\ 
 $ \rm [Ca/Fe]$&              $-0.04\pm0.35$ &\hspace{6.5pt}$ 0.03\pm0.31$ &\hspace{6.5pt}$ 0.07\pm0.31 $ &              $-0.04\pm0.32$ &              $-0.17\pm0.33$ &              $-0.03\pm 0.14$ \\ 
 $ \rm [Sc/Fe]$&              $-0.17\pm0.41$ &              $-0.19\pm0.34$ &              $-0.16\pm0.29 $ &              $-0.21\pm0.34$ &              $-0.21\pm0.26$ &              $-0.19\pm 0.14$ \\ 
 $ \rm [Ti/Fe]$&              $-0.15\pm0.36$ &              $-0.20\pm0.33$ &              $-0.14\pm0.36 $ &              $-0.21\pm0.30$ &              $-0.18\pm0.27$ &              $-0.18\pm 0.14$ \\ 
 $ \rm [V/Fe] $&              $-0.21\pm0.36$ &              $-0.17\pm0.27$ &              $-0.14\pm0.28 $ &              $-0.24\pm0.32$ &              $-0.19\pm0.28$ &              $-0.19\pm 0.13$ \\ 
 $ \rm [Cr/Fe]$&              $-0.06\pm0.30$ &              $-0.05\pm0.27$ &              $-0.07\pm0.28 $ &              $-0.06\pm0.33$ &              $-0.06\pm0.30$ &              $-0.06\pm 0.13$ \\ 
 $ \rm [Mn/Fe]$&              $-0.15\pm0.38$ &              $-0.10\pm0.30$ &              $-0.12\pm0.36 $ &              $-0.08\pm0.33$ &              $-0.08\pm0.30$ &              $-0.10\pm 0.15$ \\ 
 $ \rm [Co/Fe]$&              $-0.12\pm0.31$ &              $-0.16\pm0.28$ &              $-0.11\pm0.32 $ &              $-0.17\pm0.35$ &              $-0.14\pm0.32$ &              $-0.14\pm 0.14$ \\ 
 $ \rm [Ni/Fe]$&              $-0.09\pm0.32$ &              $-0.07\pm0.28$ &              $-0.10\pm0.32 $ &              $-0.04\pm0.32$ &              $-0.09\pm0.28$ &              $-0.08\pm 0.14$ \\ 
 $ \rm [Cu/Fe]$&\hspace{6.5pt}$ 0.08\pm0.32$ &\hspace{6.5pt}$ 0.12       $ &\hspace{6.5pt}$ 0.10        $ &\hspace{6.5pt}$ 0.09\pm0.43$ &\hspace{6.5pt}$ 0.12\pm0.59$ &\hspace{6.5pt}$ 0.10\pm 0.02$ \\ 
 $ \rm [Zn/Fe]$&\hspace{6.5pt}$ 0.19       $ &\hspace{6.5pt}\textemdash    &              $-0.26        $ &\hspace{6.5pt}$ 0.12       $ &\hspace{6.5pt}\textemdash    &\hspace{6.5pt}$ 0.02\pm 0.20$ \\ 
 $ \rm [Sr/Fe]$&\hspace{6.5pt}$ 0.64       $ &\hspace{6.5pt}\textemdash    &\hspace{6.5pt}$ 0.46\pm0.27 $ &\hspace{6.5pt}$ 0.37\pm0.40$ &\hspace{6.5pt}$ 0.57       $ &\hspace{6.5pt}$ 0.51\pm 0.10$ \\ 
 $ \rm [Y/Fe] $&\hspace{6.5pt}$ 0.15\pm0.29$ &\hspace{6.5pt}$ 0.03\pm0.25$ &\hspace{6.5pt}$ 0.02\pm0.28 $ &\hspace{6.5pt}$ 0.04\pm0.34$ &\hspace{6.5pt}$ 0.03\pm0.27$ &\hspace{6.5pt}$ 0.05\pm 0.13$ \\ 
 $ \rm [Zr/Fe]$&\hspace{6.5pt}$ 0.05\pm0.33$ &              $-0.45\pm0.30$ &              $-0.02\pm0.43 $ &              $-0.39\pm0.41$ &              $-0.23\pm0.33$ &              $-0.22\pm 0.16$ \\ 
 $ \rm [Nb/Fe]$&\hspace{6.5pt}$ 0.22       $ &\hspace{6.5pt}\textemdash    &\hspace{6.5pt}$ 0.20        $ &\hspace{6.5pt}\textemdash    &\hspace{6.5pt}\textemdash    &\hspace{6.5pt}$ 0.21\pm 0.01$ \\ 
 $ \rm [Mo/Fe]$&\hspace{6.5pt}$ 0.13       $ &\hspace{6.5pt}$ 0.20       $ &\hspace{6.5pt}$ 0.19        $ &\hspace{6.5pt}$ 0.12       $ &\hspace{6.5pt}$ 0.17       $ &\hspace{6.5pt}$ 0.16\pm 0.03$ \\ 
 $ \rm [Ru/Fe]$&\hspace{6.5pt}\textemdash    &              $-0.22       $ &\hspace{6.5pt}\textemdash     &              $-0.15       $ &              $-0.06       $ &              $-0.14\pm 0.07$ \\ 
 $ \rm [Ba/Fe]$&\hspace{6.5pt}$ 0.71\pm0.32$ &\hspace{6.5pt}$ 0.72       $ &\hspace{6.5pt}$ 0.69        $ &\hspace{6.5pt}$ 0.75\pm0.24$ &\hspace{6.5pt}$ 0.67       $ &\hspace{6.5pt}$ 0.71\pm 0.03$ \\ 
 $ \rm [La/Fe]$&\hspace{6.5pt}$ 0.27\pm0.31$ &\hspace{6.5pt}$ 0.16\pm0.22$ &\hspace{6.5pt}$ 0.32\pm0.30 $ &\hspace{6.5pt}$ 0.22\pm0.38$ &\hspace{6.5pt}$ 0.15\pm0.26$ &\hspace{6.5pt}$ 0.21\pm 0.13$ \\ 
 $ \rm [Ce/Fe]$&\hspace{6.5pt}$ 0.25\pm0.40$ &\hspace{6.5pt}$ 0.55\pm0.40$ &\hspace{6.5pt}$ 0.13\pm0.41 $ &\hspace{6.5pt}$ 0.50\pm0.35$ &\hspace{6.5pt}$ 0.26\pm0.29$ &\hspace{6.5pt}$ 0.34\pm 0.16$ \\ 
 $ \rm [Pr/Fe]$&\hspace{6.5pt}$ 0.27\pm0.53$ &\hspace{6.5pt}$ 0.23\pm0.24$ &\hspace{6.5pt}\textemdash     &\hspace{6.5pt}$ 0.06\pm0.44$ &              $-0.23\pm0.28$ &\hspace{6.5pt}$ 0.08\pm 0.20$ \\ 
 $ \rm [Nd/Fe]$&\hspace{6.5pt}$ 0.09\pm0.38$ &\hspace{6.5pt}$ 0.04\pm0.32$ &\hspace{6.5pt}$ 0.07\pm0.32 $ &\hspace{6.5pt}$ 0.15\pm0.32$ &\hspace{6.5pt}$ 0.06\pm0.30$ &\hspace{6.5pt}$ 0.08\pm 0.15$ \\ 
 $ \rm [Sm/Fe]$&\hspace{6.5pt}$ 0.81\pm0.67$ &\hspace{6.5pt}\textemdash    &\hspace{6.5pt}$ 0.73\pm0.40 $ &\hspace{6.5pt}$ 0.63\pm0.32$ &\hspace{6.5pt}$ 0.44       $ &\hspace{6.5pt}$ 0.65\pm 0.14$ \\ 
 $ \rm [Gd/Fe]$&              $-0.03       $ &\hspace{6.5pt}\textemdash    &\hspace{6.5pt}\textemdash     &\hspace{6.5pt}\textemdash    &\hspace{6.5pt}\textemdash    &              $-0.03        $ \\ 
 $ \rm [Dy/Fe]$&              $-0.08       $ &\hspace{6.5pt}\textemdash    &\hspace{6.5pt}$ 0.08        $ &\hspace{6.5pt}\textemdash    &\hspace{6.5pt}\textemdash    &\hspace{6.5pt}$ 0.00\pm 0.08$ \\ 
\hline                                                                         
\end{tabular}                                                                  
\end{center}                                                                   
\end{table*}

The chemical abundances were determined by means of the spectrum synthesis method and the code \textsc{FITSUN} by using the '$\log g$ free' 
atmospheric parameters computed with \textsc{FITSUN} and provided in Table~\ref{table:atmos:giants}. The code \textsc{SME} has not been used to perform separate
computations because (1) both \textsc{FITSUN} and \textsc{SME} adopt the same physical description of the radiative transfer, i.e. 1D\,LTE (the model atmospheres 
used in this paper are LTE and the NLTE abundances obtained with \textsc{SME} from Fe lines are computed by applying NLTE corrections), (2) the atmospheric 
parameters of our targets derived with both codes which might change the element abundances are very similar which implicate similarity of abundances computed 
with the two codes. Finally, we note that although computing NLTE abundances for elements other than iron is feasible, it is beyond the scope of this paper. 

We derived the abundances of 31 elements by using all available spectral features including isolated and blended lines. The number of elements 
for which we derived abundances differs from one star to another varying between 25~and 30~because the respective features were not of sufficient quality in all 
spectra. The results are provided in Table~\ref{elements:abs} which lists the absolute values of chemical abundances, $\log \epsilon(X) = \log (N_X/N_H) + 12$, 
their standard deviations, and, in square brackets, the number of lines used in the analysis. The uncertainties are standard deviations of abundances derived from 
individual spectral features detected for a given element. For those elements for which we detected only one or two features, standard deviations were not computed. 
The solar abundances by \citet{grevesse98} which were used in these computations are listed in the last column of the table. 

In case of iron, in Table~\ref{elements:abs}, we give the abundances computed by using three approaches. In the first one, which is consistent with 
the way the abundances of the remaining elements were computed, we used all available spectral features of Fe~I and Fe~II, blended and unblended. Then, we provide 
the iron abundances obtained by using only unblended isolated lines of Fe~I, which is consistent with the atmospheric parameters provided in 
Table~\ref{table:atmos:giants}, and finally the abundances obtained by using only unblended isolated lines of Fe~II. All these values agree well to within $1\sigma$ 
of their error bars.

Table~\ref{elements:rel} provides the ratios of the abundances of elements with respect to the abundance of iron ($[X\rm /Fe]$) calculated using the values in 
Table~\ref{elements:abs} (in case of iron we used the values obtained from Fe~I lines only) according to the formula:
\begin{equation}
\label{abun_eq}
\bigg[\frac{X}{\rm Fe}\bigg]_{\mbox {\ding{73}} } = \bigg[\frac{X}{\rm H}\bigg]_{\mbox {\ding{73}} } - \bigg[\frac{\rm Fe}{\rm H}\bigg]_{\mbox {\ding{73}} }
    \mbox{, where}
\end{equation}
\begin{equation}
    \bigg[\frac{X}{\rm H}\bigg]_{\mbox {\ding{73}} } = \log \bigg(\frac{N_{X}}{N_{\rm H}}\bigg)_{\mbox {\ding{73}} } - \log \bigg(\frac{N_{X}}{N_{\rm H}}\bigg)_{\sun},
\end{equation}
where the symbols '$\mbox {\ding{73}}$' and '$\sun$' refer to the programme star and the Sun, respectively. Like in Table~\ref{elements:abs}, we provide in
Table~\ref{elements:rel} the standard deviations only for those elements, for which at least three spectral features were detected.

The standard deviations given in Tables~\ref{elements:abs} and~\ref{elements:rel} do not include systematic differences which are due to the choice of the 
atomic data, the adopted model atmosphere, the reference of the solar abundances, the algorithm used for choosing the best fit to spectrum, the ambiguities 
related with the placement of the continuum, and the impact of the adopted stellar atmospheric parameters $(T_{\rm eff}, \log g, \mbox{and } \xi_{\rm t})$. 
In order to estimate the last contribution, we altered the values of $T_{\rm eff}$, $\log g$, and $\xi_{\rm t}$ by, respectively, 50~K, 0.1~dex, and 
0.5~km\,s$^{-1}$. The uncertainties of [Fe/H] resulting from those variations are listed in Table~\ref{stel-par-fe}. 

\begin{figure}
\includegraphics[width=8.5cm,angle=0]{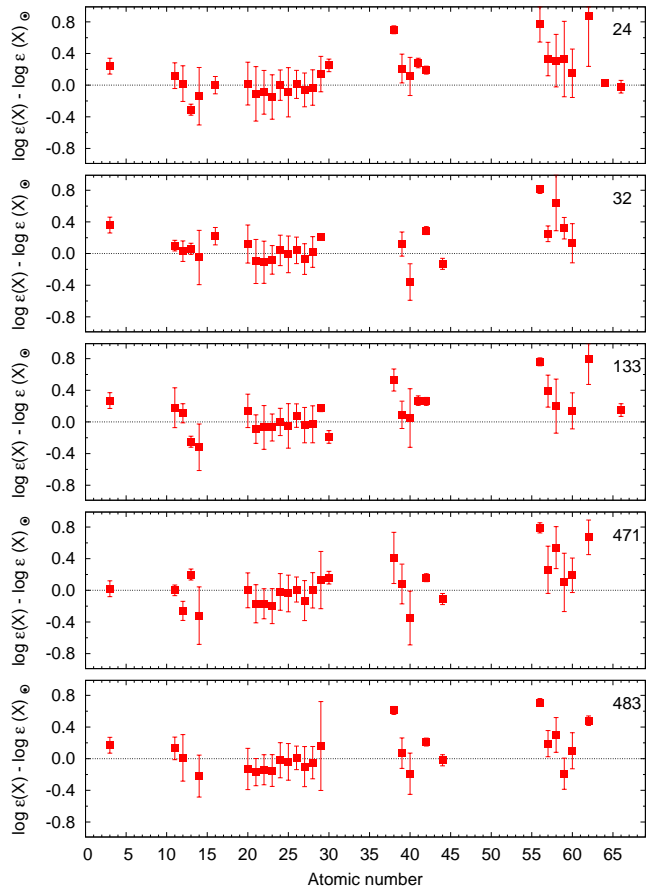}
\caption[]{Differences between the chemical abundances of the G-type giants~24, 23, 133, 471, and 483, and the Sun.}
\label{figure:Gtype-Sun}
\end{figure}

\begin{table}
\begin{center}
    \caption{The changes of $\rm [Fe/H]$ resulting from the changes of $T_{\rm eff}$, $\log g$, and $v_{\rm t}$.}
\label{stel-par-fe}
\begin{tabular}{rccc}
\hline\hline\noalign{\smallskip}
WEBDA (KIC) &$\Delta T_{\rm eff}$ & $\Delta \log g$ & $\Delta v_{\rm t}$ \\
            &         $\pm 50$~K&   $\pm 0.1$~dex   &   $\pm 0.5$~km\,s$^{-1}$\\
\hline
 24 (9655101)& $\pm 0.039$ &$\pm 0.008$ &$\mp 0.141$\\  
 32 (9655167)& $\pm 0.035$ &$\pm 0.006$ &$\mp 0.162$\\  
133 (9716090)& $\pm 0.030$ &$\pm 0.013$ &$\mp 0.095$\\  
471 (9776739)& $\pm 0.030$ &$\pm 0.009$ &$\mp 0.144$\\  
483 (9532903)& $\pm 0.027$ &$\pm 0.018$ &$\mp 0.121$\\  
\hline
\end{tabular}
\end{center}
\end{table}

The general pattern of abundances in our programme stars is consistent with solar values by \citet{grevesse98} as shown in 
Fig.~\ref{figure:Gtype-Sun}. Slightly higher differences between abundances derived in this paper, and the solar values can be noticed for a few rare-earth 
elements\footnote{According to the definition of the International Union of Pure and Applied Chemistry (IUPAC), rare earth elements consist of~15 lanthanides 
(La, Ce, Pr, Nd, Pm, Sm, Eu, Gd, Tb, Dy, Ho, Er, Tm, Yb, and Lu) plus scandium (Sc) and yttrium (Y).} for which only few lines were available for analysis, 
and for Ba which we find overabundant and which we discuss in a next paragraph.

In the last column of Table~\ref{elements:rel}, we provide the weighted mean ratios of chemical elements to iron for the whole cluster calculated from the 
individual values measured for the five G-type stars. If all individual measurements of an element has an uncertainty given in Table~\ref{elements:rel}, then
the column 'cluster mean' provides a weighted mean and uncertainty. If, for a given element, any star in Table~\ref{elements:rel} had a measurement without an 
uncertainty, we used a flat mean and the star-to-star RMS scatter as the uncertainty. In the case of Gd where there was only one measurement, we do not provide 
an uncertainty for the cluster mean.

Most of these values are very close to solar, however, there are few exceptions. We comment upon some of them in the following, cautioning the reader to keep 
in mind the difficulty of comparing results from different spectroscopic studies. One of those exceptions is a very high abundance of barium (the mean value for 
the cluster $\rm [Ba/Fe] = 0.71\pm0.03$) which may be expected in the youngest clusters but not in the oldest \citep[see][]{dorazi2009}. Indeed, there is very
few open clusters of an age similar to NGC\,6811 and a comparably high abundance of barium: NGC\,2324, $age = 0.67$~Gyr \citep{salaris2004}, $\rm [Ba/Fe] = 
0.66\pm0.09$ \citep{dorazi2009} and an even older cluster NGC\,2141, $age = 2.45$~Gyr \citep{salaris2004}, $\rm [Ba/Fe] =0.91$ derived by \cite{yong2005} 
\citep[but see][who obtained a value lower by a half]{jacobson2013}. Therefore, we conclude that even though the Ba abundance which we report is high, it is still 
consistent with other open clusters of an age near 1~Gyr.

\begin{figure*}
\includegraphics[width=5.5cm,angle=0]{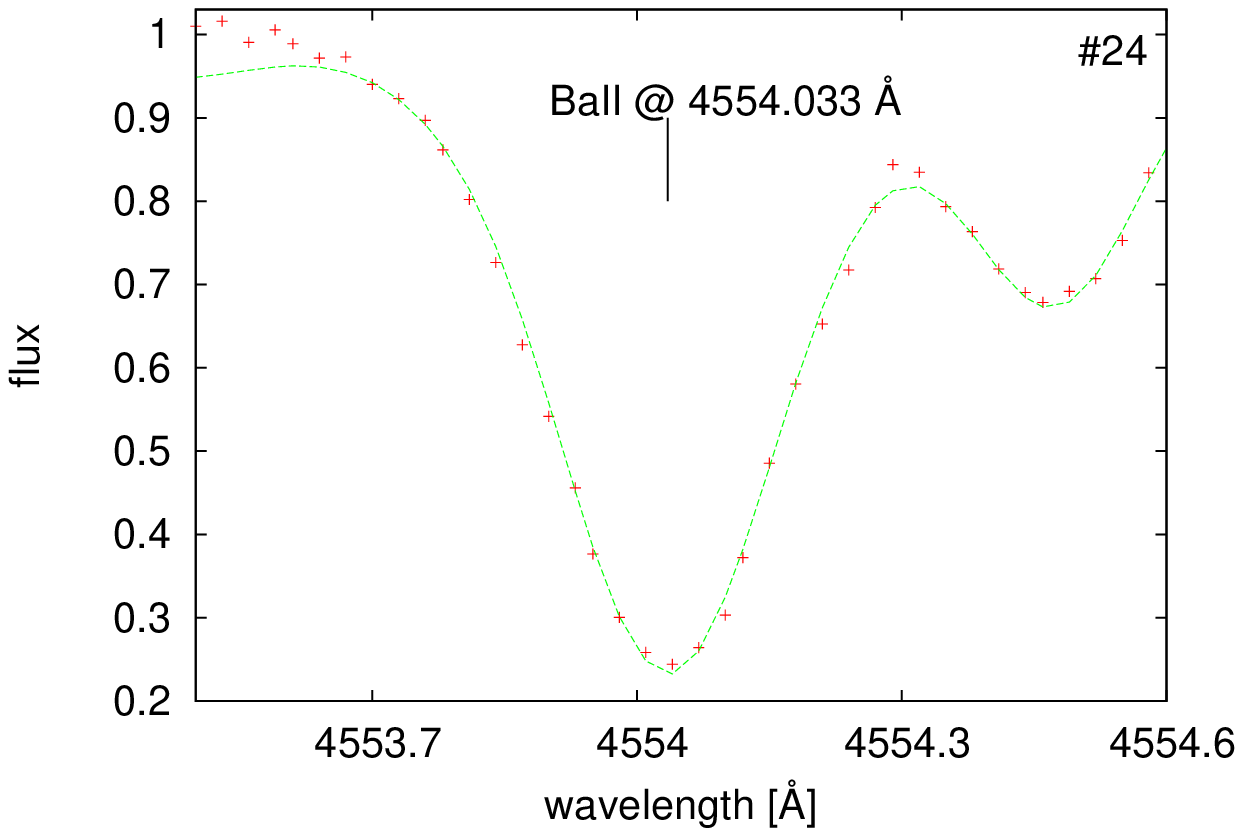}
\includegraphics[width=5.5cm,angle=0]{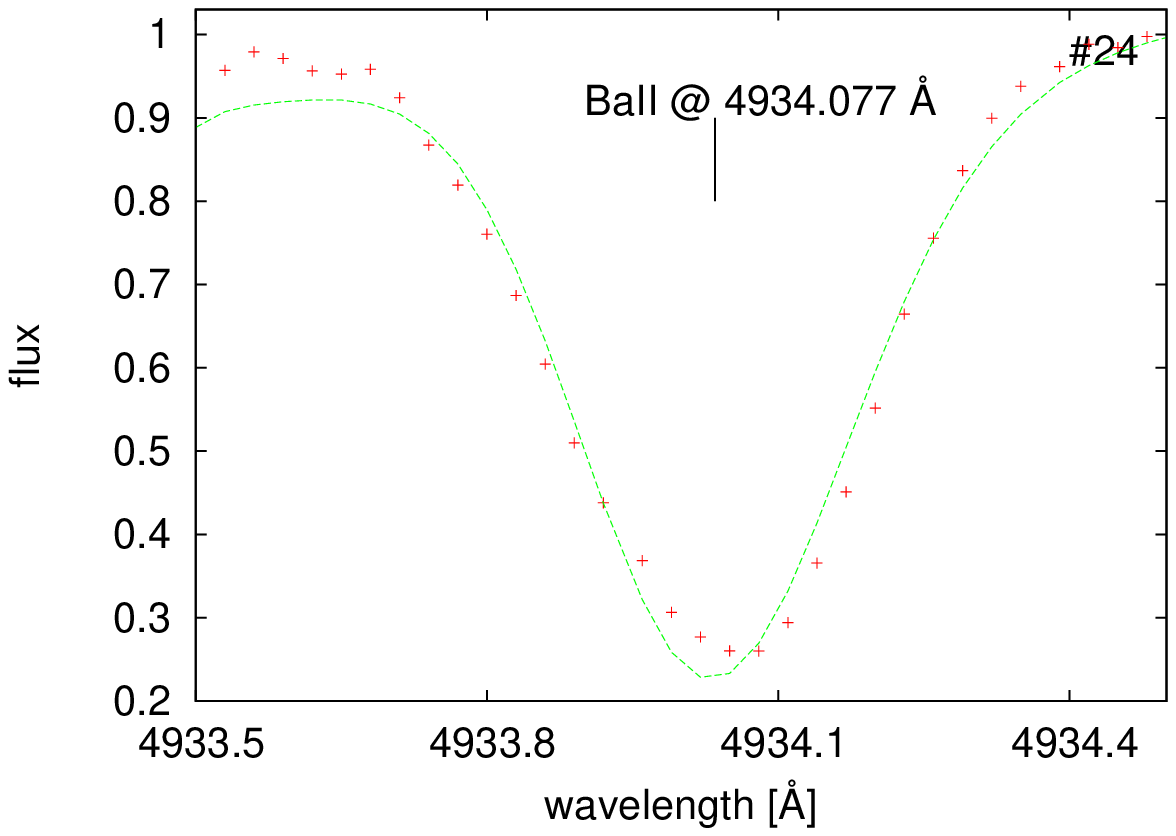}
\includegraphics[width=5.5cm,angle=0]{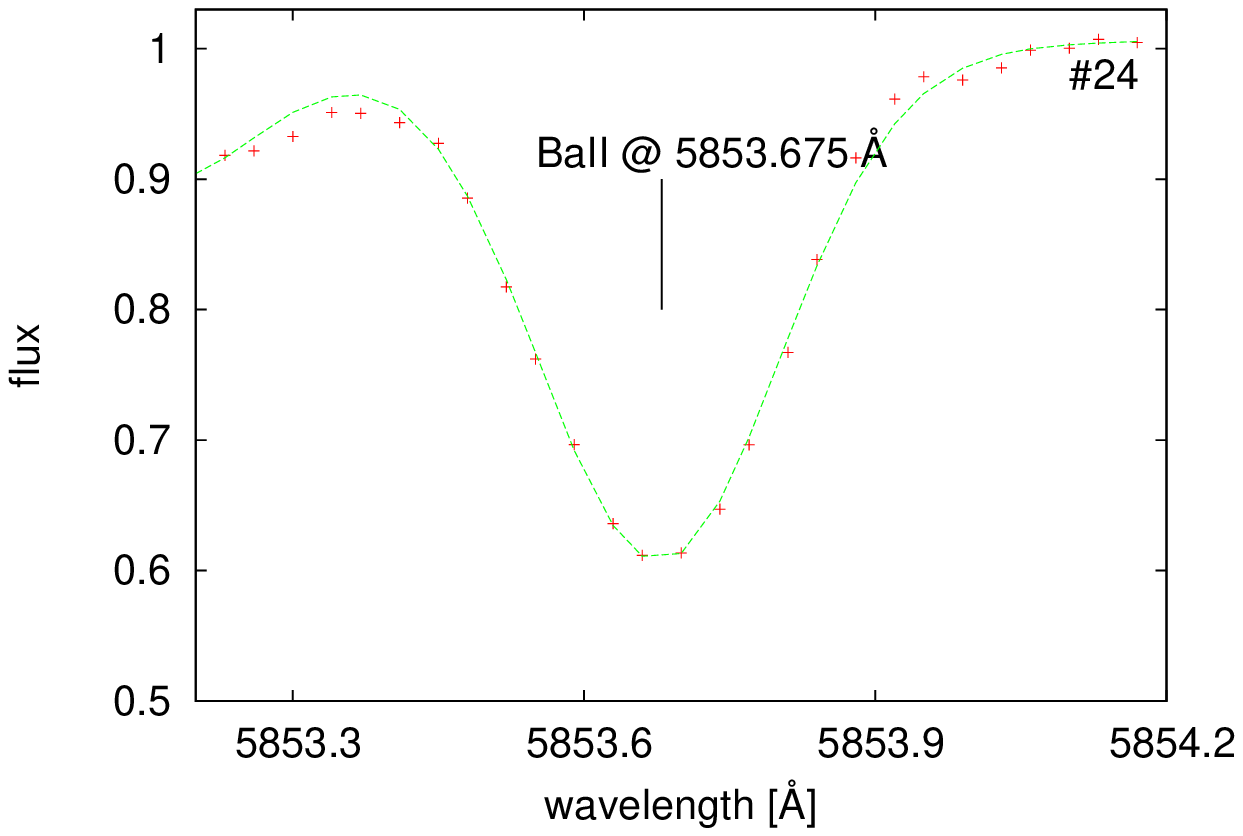}
\includegraphics[width=5.5cm,angle=0]{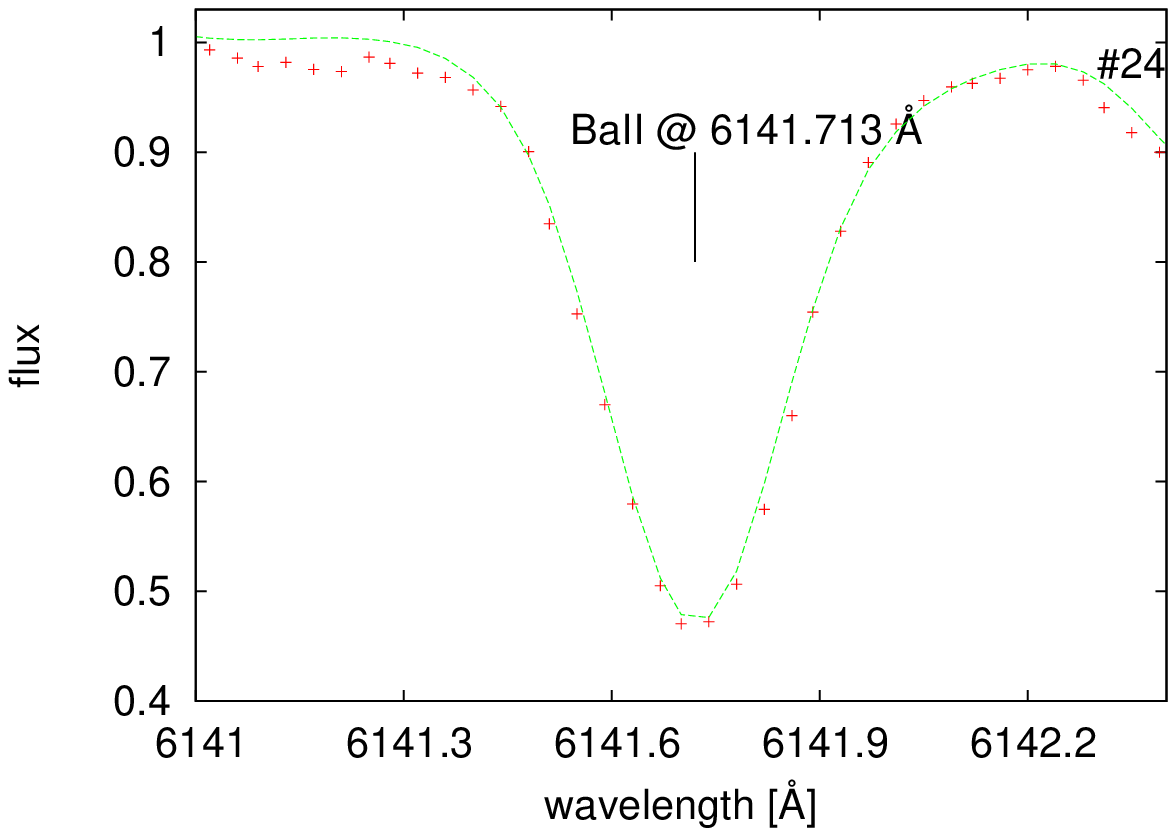}
\includegraphics[width=5.5cm,angle=0]{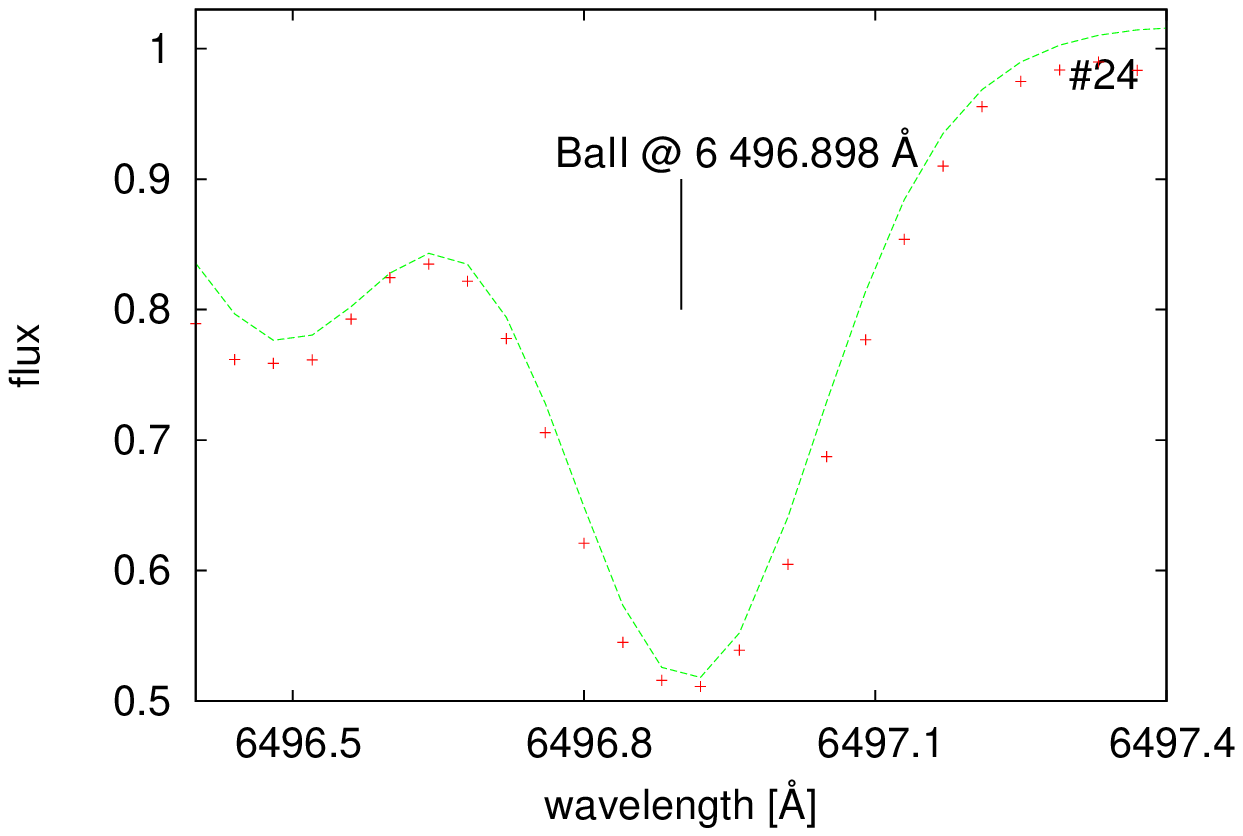}
\includegraphics[width=5.5cm,angle=0]{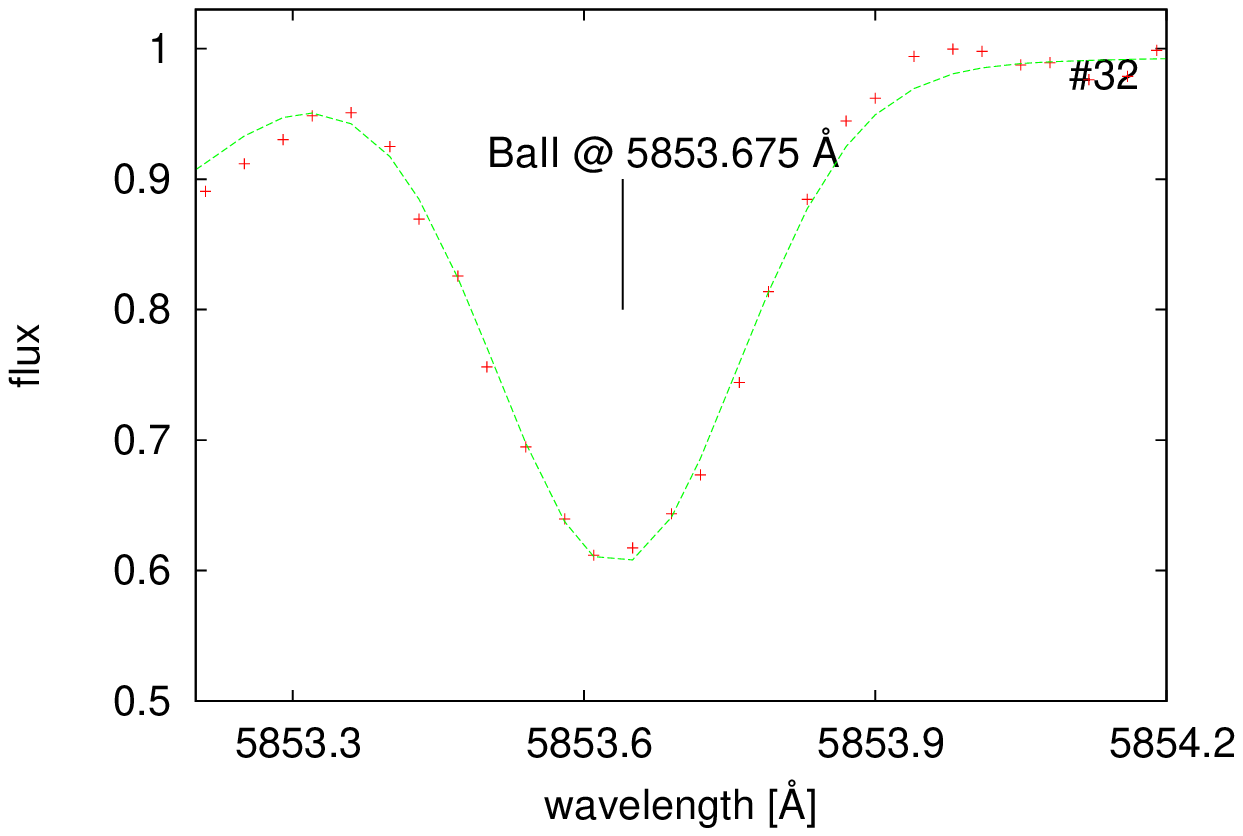}
\includegraphics[width=5.5cm,angle=0]{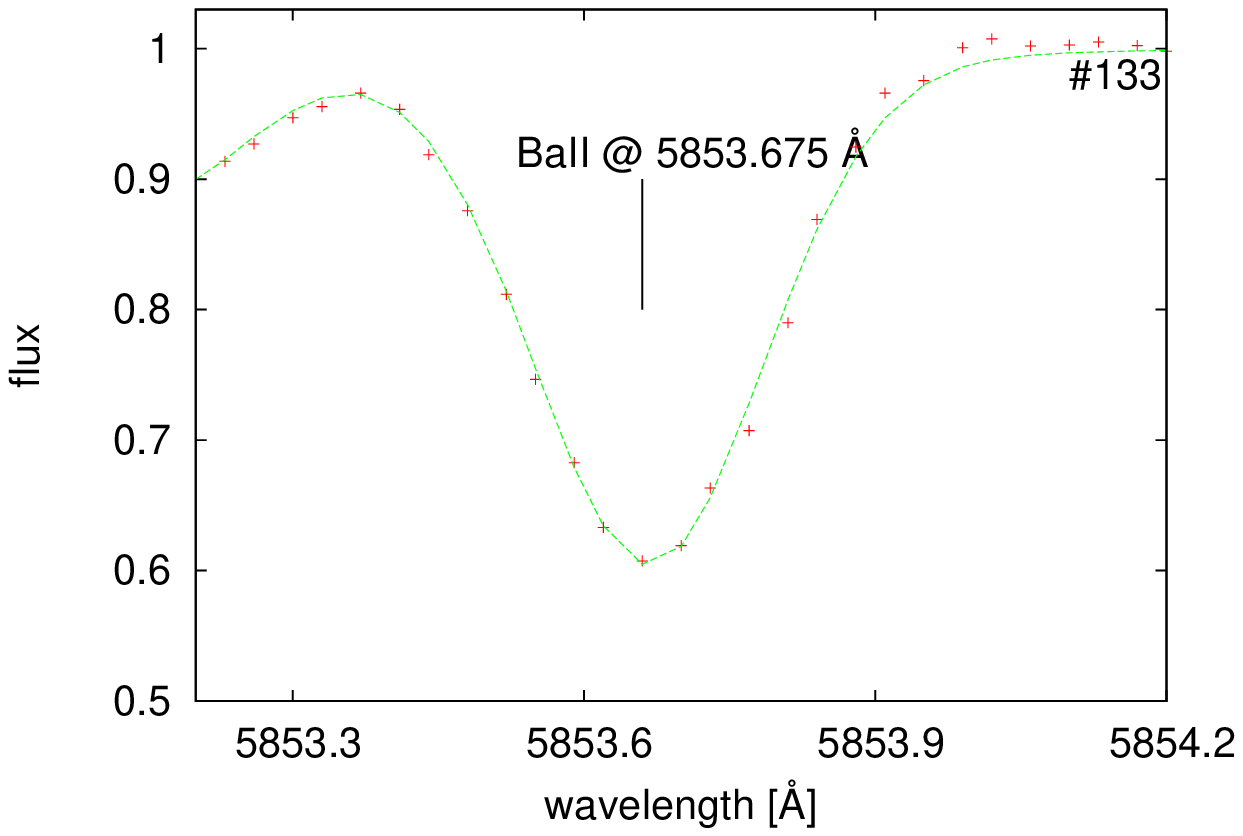}
\includegraphics[width=5.5cm,angle=0]{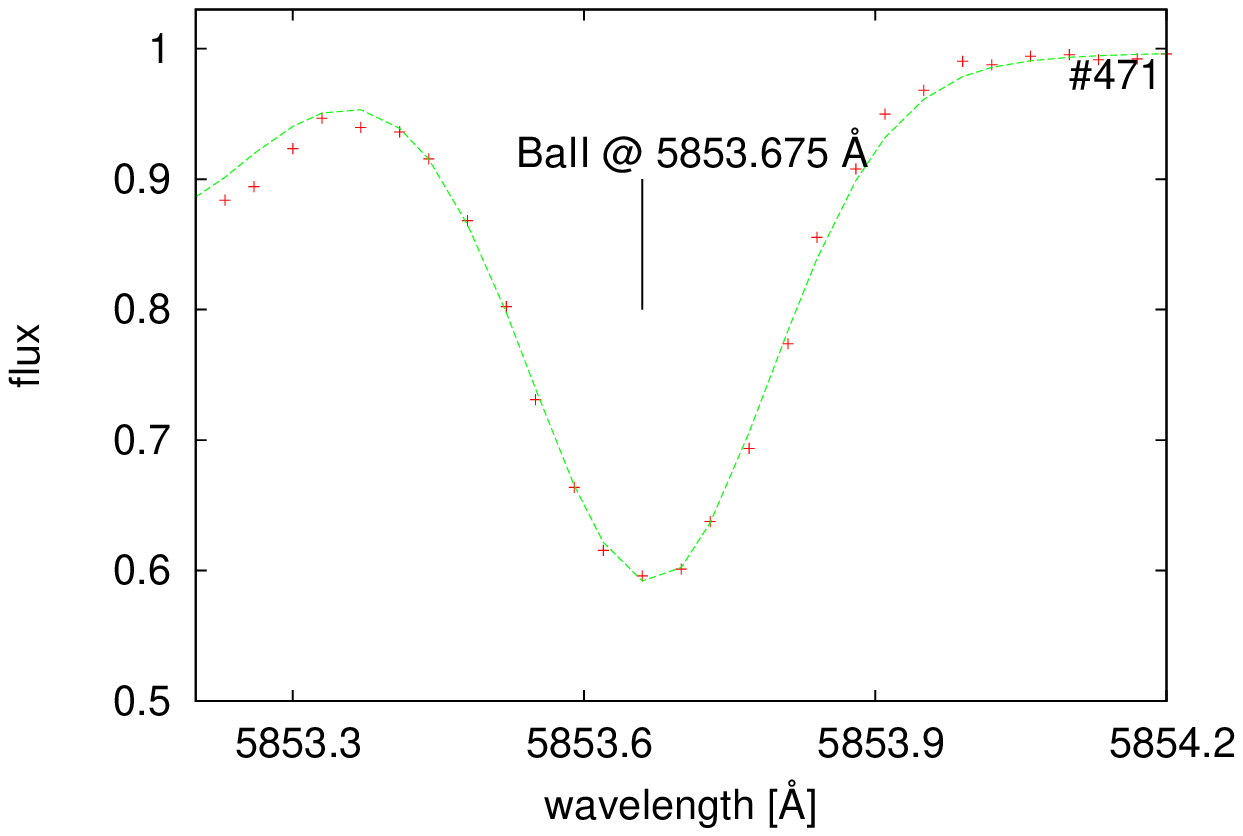}
\includegraphics[width=5.5cm,angle=0]{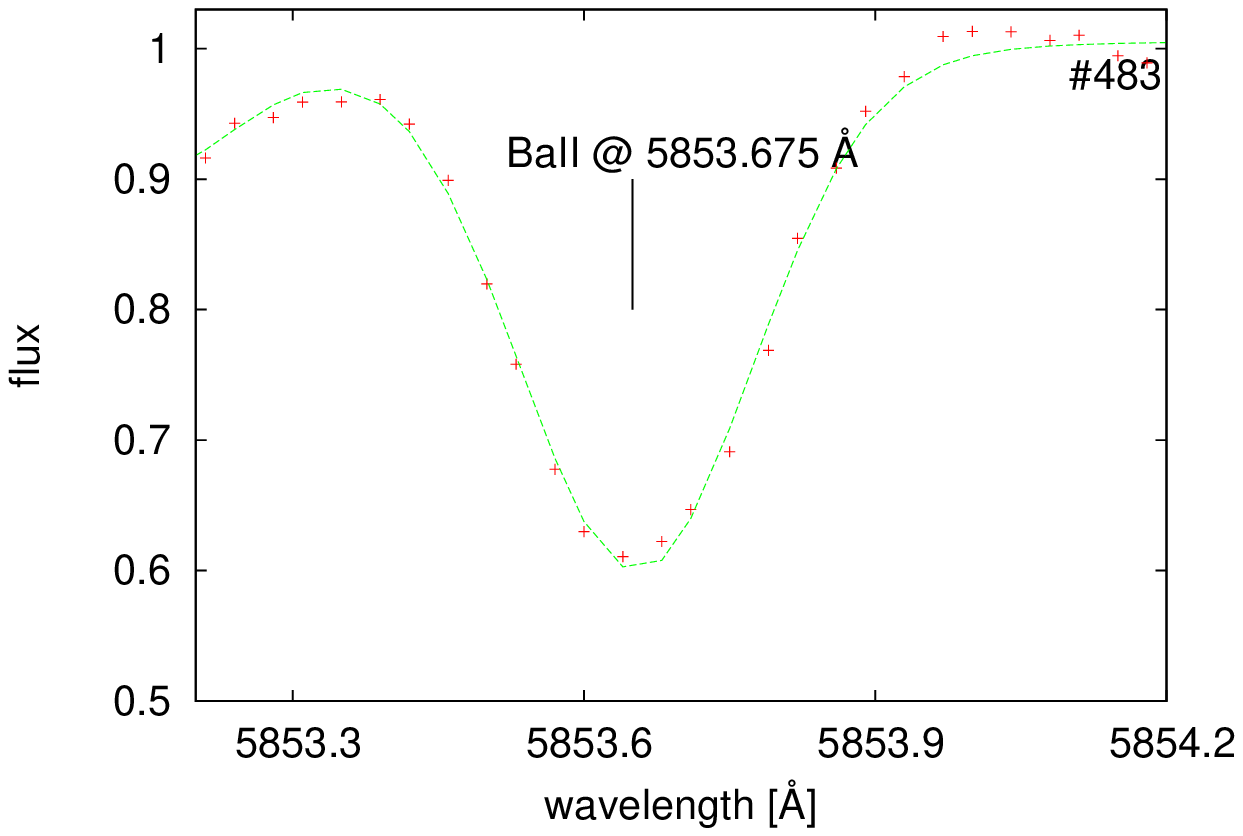}
\caption[]{The synthetic and the observed spectra of stars~24, 32, 133, 471, and 483, centred at the lines of Ba analysed in this paper. For star~24, we show
all the Ba lines used in the analysis. For the remaining stars, only the Ba\,II line at 5853.675 \AA\, is shown. The observed spectra are plotted with crosses 
while the synthetic ones are plotted with a solid line. The position of Ba lines are indicated. The star numbers are indicated in the right top corner of each 
panel.}
\label{ba_flux}
\end{figure*}

In Fig.~\ref{ba_flux} we plot the synthetic and the observed spectra of stars~24, 32, 133, 471, and 483, centred at the lines of barium which have been analysed 
by us. For star~24, we show all the lines of Ba which were used in our analysis. All those lines yielded consistent abundance of barium. That concerns also the 
line of Ba\,II at 4934.077\,\AA\, which shows slight asymmetry due to difficulties in the continuum placement. For the other stars, where fewer lines were 
measured, we show only the fit of a Ba\,II line at 5853.675\AA. Like for the other elements, the abundance of Ba has been derived from all available spectral 
features. Our computations included the hyperfine splitting.

Another interesting element is zirconium. Our ratio of $\rm [Zr/Fe] = -0.22\pm0.16$ is lower than expected for a 1~Gyr-old open cluster according to figure~9 of 
\citet{maiorca2011} and as such is more consistent with the conclusion of \citet{jacobson2013} that [Zr/Fe] shows no trend with age. Finally, we report that the 
abundance of Li in all the five G-type stars has been found to be higher than expected for red clump stars \citep[c.f., e.g.,][]{villanova2010, gonzalez2009}. 
We note, however, that the respective spectral region at 6707.76~{\AA} and 6707.91~{\AA} where the spectral features of Li are located was found to be useless in 
the data acquired on 28 June 2007 for all stars but~24. Therefore, our determination of the abundance of lithium in red clump stars in NGC\,6811 should be taken 
with caution.

\begin{table*}
\caption{Properties of red clump stars in NGC~6811.}
\label{table:properties}    
\begin{tabular}{p{5.0cm}p{1.5cm}p{1.5cm}p{1.5cm}p{1.5cm}p{1.5cm}p{1.5cm}}
\hline\noalign{\smallskip}
Property & \multicolumn{5}{c}{WEBDA (KIC)} & Average\\
& 24 (9655101) & 32 (9655167) & 133 (9716090) & 471 (9776739) & 483 (9532903) \\
\noalign{\smallskip}\hline\noalign{\smallskip}
$\nu _{\rm{max}}$ [$\mu$Hz]	& $98.2\pm2.4$   & $100.3\pm8.7$  & $101.4\pm5.9$  & $93.4\pm9.0$    & $96.3\pm4.5$   & --\\
$\Delta \nu$  [$\mu$Hz]		& $7.86\pm0.04$  & $8.07\pm0.04$  & $8.56\pm0.06$  & $7.93\pm0.16$   & $7.69\pm0.16$  & --\\
$T_{\rm{eff}}$ [K]     		& $5005\pm100$   & $4924\pm100$   & $4980\pm100$   & $4952\pm100$    & $5008\pm100$   & --\\ 
$V$                         & 11.236         & 11.318         & 11.372         & 11.159 & 11.172 & --\\
\noalign{\smallskip}\hline\noalign{\smallskip}
From asteroseismic scaling relations:\\
$M/M_{\odot}$   & $ 2.23\pm0.18$ & $ 2.09\pm0.66$ & $ 1.73\pm0.33$ & $ 1.82\pm0.60$ & $ 2.30\pm0.39$ & $2.12\pm0.14$\\
$R/R_{\odot}$   & $ 8.70\pm0.25$ & $ 8.36\pm0.74$ & $ 7.55\pm0.46$ & $ 8.08\pm0.85$ & $ 8.91\pm0.58$ & $8.47\pm0.19$\\
$(m-M)_{\rm v}$ & $10.32\pm0.15$ & $10.22\pm0.24$ & $10.12\pm0.19$ & $10.02\pm0.27$ & $10.32\pm0.19$ & $10.23\pm0.08$ \\
\noalign{\smallskip}\hline\noalign{\smallskip}
From alternative approach:\\
$\delta\nu _{\rm{max}}$ [$\mu$Hz] &  1.00  &  3.30  & 10.05  &  7.52  &  0.03  & --    \\
$M/M_\odot$ (fixed)		          &  2.30 &  2.30 &  2.30 &  2.30 &  2.30 &  2.30 \\
$R/R_\odot$                       &  8.79 &  8.63 &  8.30 &  8.74 &  8.92 & 8.68    \\ 
$(m-M)_{\rm v}$ 		          & 10.34 & 10.29 & 10.33 & 10.19 & 10.32 & 10.29 \\
\noalign{\smallskip}\hline\noalign{\smallskip}
\end{tabular}
\end{table*}

\section{Cluster parameters}
\label{cluster_param}

With our spectroscopic measurements of metallicity and effective temperatures, a detailed asteroseismic investigation of the cluster stars can now begin. 
Although this is beyond the scope of the present paper, we can already make some inferences on the cluster properties by utilising our spectroscopic 
measurements and the asteroseismic measurements of the helium burning red clump (RC) stars available at this point. We calculated the masses, radii and 
apparent distance moduli from the asteroseismic scaling relations in the form of eqns.~(3) and~(4) of \cite{miglio2012}. The input values for these calculations 
are given in the top part, and results in the middle part of Table~\ref{table:properties}. We adopted the $T_{\rm{eff}}$ values derived with the code 
\textsc{SME} with $\log g$ fixed at the asteroseismic value, and an uncertainty of $\pm$100~K consistent with the biases reported in Table~\ref{table:biases}. 
The choice of using results from \textsc{SME} was based mainly on the low star-to-star scatter in $\rm [Fe/H]$ values obtained from that code. 
We note however that the cluster parameters derived in this section would only be affected at a level below the $1\,\sigma$  uncertainties by adopting the 
\textsc{FITSUN} values instead. The abundances of other elements than Fe were only measured with \textsc{FITSUN} and we adopt the scaled solar result. Since 
these are measurements relative to Fe, this result will not be significantly affected by differences in the stellar parameters at the level we find between 
methods. The $V$-band magnitudes used in these computations were adopted from \citet{glushkova1999}.

Due to rather large uncertainties, on $\nu_{\rm max}$ in particular, the masses of the RC stars are not as precisely determined as we would like when calculated 
from the asteroseismic scaling relations. Therefore, we also adopted the alternative approach of estimating the asteroseismic RC masses from figure~4 of 
\cite{stello2013}. In that figure, valid for solar metallicity, it can be seen that the maximum value of $\Delta \nu$ for a helium core-burning star 
defines the mass of the star, regardless of the $\Delta P$ value. From the maximum value of $\Delta \nu$ for our stars in Table~\ref{table:seismic} 
(8.56~$\mu$Hz), we obtain a RC mass of 2.3~$M_{\odot}$. 

When comparing this number to the individual mass measurements done using the scaling relations, we find that they are in agreement within the uncertainties. For 
the latter determinations, we find that the errors in mass and apparent distance modulus are correlated along a line defined by uncertainties in $\nu_{\rm max}$ which has 
also been found by \citet{brogaard2014} for the open cluster NGC~6819. We only expect a very small systematic correction to $\Delta \nu$ in the case of the RC 
stars with mass and $T_{\rm{eff}}$ appropriate for NGC~6811 \citep[see figure~2 in][]{miglio2013}. Therefore, since the main source of the random uncertainty of 
mass comes from $\nu_{\rm max}$, we slightly adjusted the values of $\nu_{\rm max}$ for each star until the masses from the scaling relations resulted in 
2.3~$M_{\odot}$, consistent with our alternative approach. These corrections ($\delta \nu_{\rm max}$) were always positive but much less than $1\sigma$ for three stars, 
$0.8\sigma$ for one star and $1.7\sigma$ for another star. We then calculated the radii and apparent distance moduli of the stars using the corrected values of $\nu_{\rm 
max}$. By doing that, the asteroseismic distance moduli from the individual stars reach a much better agreement, as they should, since all cluster stars are expected 
to be at the same distance. The numbers from this alternative approach are given in the bottom part of Table~\ref{table:properties}.

\begin{figure*} 
\includegraphics[width=17.5cm,angle=0]{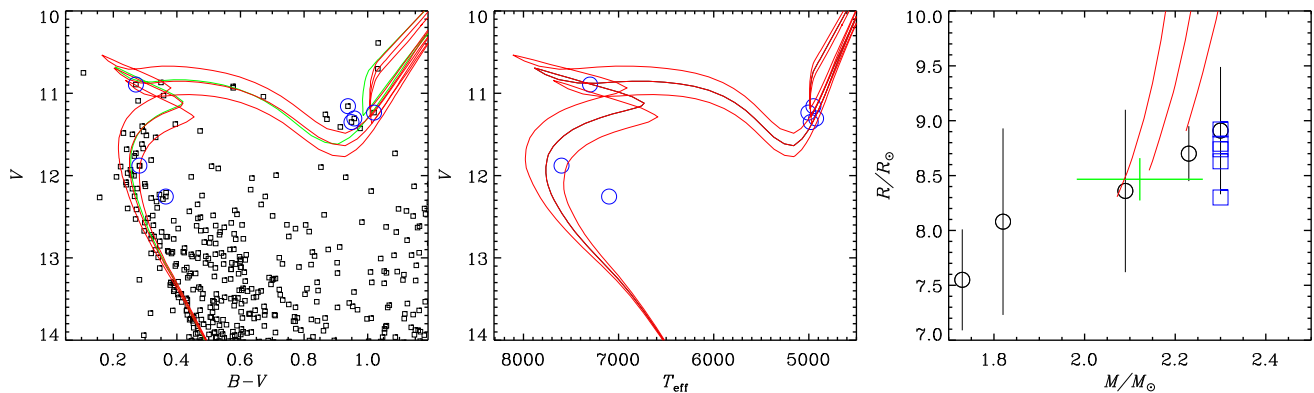}
\caption[]{Left: the $(B-V)-V$ colour-magnitude diagram of NGC~6811 from \cite{janes2013}. Blue circles indicate cluster members identified in this paper. The 
isochrones of~0.9, 1.0, and~1.1~Gyr and $\rm [Fe/H]=+0.05$ from \cite{bressan2013} are overplotted in red. The green line indicates the 1.0~Gyr isochrone transformed 
to the observational plane using \textsc{MARCS} atmosphere models instead of the colour-transformations adopted by \cite{bressan2013}. Middle: $T_{\rm eff}-V$ diagram 
of NGC~6811; blue circles indicate the same cluster members as on the left panel. Right: the masses and radii of the RC stars as determined from the scaling 
relations (black circles) and from our alternative approach (blue squares). The weighted mean value and its $1\sigma$ uncertainty is indicated by the green plus
sign. The RC stage of isochrones of~0.9, 1.0, and 1.1~Gyr with ages increasing from right to left are overplotted.}
\label{fig:cmd-karsten} 
\end{figure*}

On the left panel of Fig.~\ref{fig:cmd-karsten}, we show the $(B-V)-V$ colour-magnitude diagram (CMD) of NGC~6811 from the observations of \cite{janes2013}. Our 
RC stars and hot single cluster members with no spectroscopic evidence for binarity (stars 33, 54 and 528; Table~\ref{rv-individual-6811}) are indicated 
with blue circles, and PARSEC isochrones \citep{bressan2013} having $[Fe/H]=+0.05$ and ages of 0.9, 1.0, and 1.1~Gyr, respectively, are shown. To put the 
isochrones on the observational plane, we used an apparent distance modulus of 10.29~mag calculated from our spectroscopic $T_{\rm eff}$ measurements obtained 
with \textsc{SME} for the asteroseismic $\log g$ values (see the middle right-hand part of Table~\ref{table:atmos:giants}), the bolometric corrections from 
\textsc{MARCS} atmosphere models \citep{casagrande2014} and the asteroseismic measurements from Table \ref{table:seismic} with $\nu_{\rm max}$ slightly corrected 
as described above (see Table~\ref{table:properties}.) A reddening of $E(B-V)=0.05$ was applied to the isochrones in an attempt to match the observed main 
sequence. In order to get the best reddening estimate, we assumed that our three hot single member stars should have the same position relative to the isochrone 
on colour and $T_{\rm eff}$ planes by comparing the left and middle parts of Fig.~\ref{fig:cmd-karsten}. The faintest of the hot stars (star 54) appears to 
be an equal mass binary from its location in the left and middle panels, although we found no clear spectroscopic evidence for that. However, the star can still 
be used for our procedure, since we are only requiring the position of the star relative to the isochrone to be similar in the left and middle panels. As 
explained below, we could not use the RC stars for this even though they appear to match the isochrones better than the hot stars in the middle panel of 
Fig.~\ref{fig:cmd-karsten}.

To emphasise the effects of different colour-temperature calibrations, in the left panel of Fig.~\ref{fig:cmd-karsten}, we use a green line to plot the 1~Gyr 
isochrone translated to colours with the bolometric corrections from \textsc{MARCS} atmosphere models that were also used to calculate the apparent
distance modulus, in addition to the colours that come with the \textsc{PARSEC} isochrones which are plotted in red. As seen, the predicted colour of the RC is 
significantly different in these two cases. The remaining parts of the two 1~Gyr isochrones show very little difference. This illustrates one of the problems 
of estimating cluster reddening and other parameters by comparing observed and synthetic colours. This issue of not knowing whether colour-$T_{\rm eff}$ 
transformations predict a correct colour-difference between the main sequence and the RC can significantly bias results from isochrone fitting in the absence of 
spectroscopic constraints. We suspect that this is the major reason why \citet{janes2013} found a lower metallicity for NGC~6811 from photometry, as we have checked 
that a lower metallicity isochrone predicts a smaller separation between the unevolved main sequence and the RC. 

One might be tempted to rely on the \textsc{MARCS} transformations, since they appear to show better consistency between colour and $T_{\rm eff}$ when comparing the 
left and middle panels, but that might just be a coincidence caused by inaccuracies in stellar models coupled to the uncertainty of our measurements and the 
photometry. In any case, if we were to make use of the RC stars to estimate the reddening, we would obtain a reddening which is lower than what the main sequence 
suggests, $E(B-V)\sim0.03$ using the \textsc{MARCS} colour-transformations, or a reddening close to zero if we trust the colours that come with the \textsc{PARSEC} 
isochrones. Although the reddening seems to be rather uncertain, a much higher reddening of $E(B-V)=0.16$ suggested for the cluster by \citet{schlegel1998} is 
clearly too high, since such a high value would make it impossible to match the CMD with an isochrone given our constraints on $T_{\rm eff}$, [Fe/H], and apparent 
distance modulus. This explains why \citet{hekker2011}, who based their analysis on photometric $T_{\rm eff}$ values obtained by assuming $E(B-V)=0.16$, obtained an 
$\rm [Fe/H]$ value for the cluster that is much too low.
 
The isochrones of age 0.9, 1.0, and 1.1~Gyr match the CMD about equally well, with a slight preference towards 1~Gyr, although that depends rather strongly on the 
interpretation of the brightest blue stars, which itself depends on the amount of convective core overshoot assumed in the stellar models. Ages older than 1.1~Gyr 
and younger than 0.9~Gyr seem to be disfavoured by the CMD, although a different amount of convective core overshoot would allow a slightly larger range of ages. 

In the right panel of Fig. \ref{fig:cmd-karsten}, we show the masses and radii of the RC stars as determined from the scaling relations in black with the weighted 
mean value and its $1\sigma$ uncertainty marked by the green plus, and measurements from our alternative approach in blue. We compare those to the RC phase of the 
same isochrones as in the other panels, with ages increasing from right to left. Uncertainties on mass measurements are not shown, but they are given in 
Table~\ref{table:properties}. As can be seen, the inferred age of the cluster depends on which of the asteroseismic mass measurements we rely on but it is consistent with 
the range of ages inferred from the CMD. Although it is difficult to estimate the uncertainty on the mass measurements from our alternative approach, it seems 
most likely that the true masses of the RC stars are somewhere between the two estimates. Possible causes for their difference include potential small systematic 
corrections to the scaling relations, an underestimate of the $T_{\rm eff}$ values, and potential changes to the isochrones in fig.~4 in \citet{stello2013} 
when including convective core overshoot. 

A more detailed asteroseismic analysis should allow a more detailed analysis of the cluster and its evolved stars. Not only by making use of the much longer 
\textit{Kepler} time series data that are now available for these stars, but also by measuring the individual oscillation frequencies instead of only the basic 
asteroseismic measures. The additional information that comes from analysis of detached eclipsing binaries that have been identified in the cluster 
\citep{brogaard2014} will also provide useful constraints. When all such measurements are available, the cluster will be extremely well characterised and should 
be well suited to obtain new constraints on convective core overshoot \citep{Montalban2013}.

We summarise our derived properties of NGC6811 in Table~\ref{table:properties:summary}. The average $\rm [Fe/H]$ has been calculated from the mean values obtained
for G-type stars with the codes \textsc{FITSUN}, \textsc{SME}, and \textsc{ROTFIT} and reported in Table~\ref{table:biases}.

\begin{table}
\caption{Properties of NGC6811.}
\label{table:properties:summary}    
\begin{tabular}{p{3.0cm}p{4.1cm}}
\hline\noalign{\smallskip}
Cluster property & value \\
\noalign{\smallskip}\hline\noalign{\smallskip}
$\rm [Fe/H]$                 & $+0.04\pm0.01$~dex (internal error)\\
$\rm [M/Fe]$                 & scaled solar \\
age                          & $1.0\pm0.1$~Gyr \\
$(m-M)_{\rm V}$              & $10.29\pm0.14$~mag \\
$E(B-V)$                     & $0.05\pm0.02$~mag \\
$\langle M_{\rm RC} \rangle$ & $2.12\pm0.14\,M_\odot$ \\
\noalign{\smallskip}\hline\noalign{\smallskip}
\end{tabular}
\end{table}

\section{Summary}
\label{sum}

We presented results of spectroscopic observations of~15 stars in the open cluster NGC~6811. The analysis of radial-velocities allowed us to discover five 
spectroscopic systems. We classified stars~68 and~173 as SB2, stars~24 and~489, as SB1 candidates, and star~218, as SB1. The mean radial-velocity of the 
cluster derived in this paper is $+6.68\pm0.08$~km\,s$^{-1}$.

Eight stars have been classified as confirmed or very probable cluster members and three, to be non-members. For stars~68, 113, 218, and 489 which are either 
spectroscopic binaries or $\delta$~Sct variables, or both, membership was not assessed. 
The atmospheric parameters and the projected rotational velocity have been derived for 13 stars.

The red clump of NGC~6811 consists of at least six stars; stars~471 and~483 were classified as RC in this paper while stars~24, 32, and 133 have been 
classified as RC by \citet{mermilliod1990}. The sixth star, KIC~9534041, was classified as RC by \citet{stello2011}. For the five RC stars in our sample, 
the abundances of~34 chemical elements have been computed. The mean mass of the red clump stars in the cluster has been found to be $2.12 \ pm 0.14 M_{\odot}$

We showed that the mean metallicity of NGC~6811 is close to solar, and that the pattern of most elements in this cluster agrees with typical 
values observed for the Galactic open clusters of similar age and location. An exception is barium which we find to be overabundant.

The age of NGC~6811 has been found to be $1\pm0.1$~Gyr.

\section*{Acknowledgements}
{\small JM-\.Z and EN acknowledge the Polish MNiSW grant N\,N203\,405139. 
KB acknowledges support from the Carlsberg Foundation and the Villum foundation. 
EN acknowledges the support from 1007/S/IAs/14 funds and funding through NCN grant 2011/01/B/ST9/05448.
MB acknowledges the support from the European Union FP7 programme through ERC grant number 320360. 
The computations have been carried out in Wroc{\l}aw Centre for Networking and Supercomputing (\url{http://www.wcss.wroc.pl}), grants No.~214 and~224.  
Funding for the Stellar Astrophysics Centre is provided by The Danish National Research Foundation (Grant DNRF106).
The research is supported by the ASTERISK project (ASTERoseismic Investigations with SONG and {\it Kepler}) funded by the European Research Council 
(Grant agreement no.: 267864). We made use of the Aladin software, the SAO/NASA's Astrophysics Data System, and the WEBDA database operated at the Institute for 
Astronomy of the University of Vienna.

\label{lastpage}


\begin{thebibliography}{99}

\bibitem[\protect\citeauthoryear{Becker}{1947}]{becker1947}                                  Becker W., 1947, Astronomische Nachrichten, 275, 229
\bibitem[\protect\citeauthoryear{Bergemann}{2011}]{bergemann2011}                            Bergemann M., 2011, MNRAS, 413, 2184
\bibitem[\protect\citeauthoryear{Bergemann \& Cescutti}{2010}]{bergemann2010}                Bergemann M., Cescutti G., 2010, A\&A, 522, A9 
\bibitem[\protect\citeauthoryear{Bergemann \& Gehren}{2008}]{bergemann2008}                  Bergemann M., Gehren T., 2008, A\&A, 492, 823 
\bibitem[\protect\citeauthoryear{Bergemann et~al.}{2012a}]{bergemann2012a}                   Bergemann M., Hansen C.~J., Bautista M., Ruchti G., 2012, A\&A, 546, A90 
\bibitem[\protect\citeauthoryear{Bergemann et~al.}{2012b}]{bergemann2012b}                   Bergemann M., Kudritzki R.-P., Plez B., Davies B., Lind~K., Gazak Z., 2012, ApJ, 751, 156 
\bibitem[\protect\citeauthoryear{Bergemann et~al.}{2012c}]{bergemann2012c}                   Bergemann M., Lind K., Collet R., Magic Z., Asplund M., 2012, MNRAS, 427, 27
\bibitem[\protect\citeauthoryear{Borucki et~al.}{2003}]{borucki2003}                         Borucki W.~J. et~al., 2003, in Society of Photo-Optical Instrumentation Engineers (SPIE) Conference Series, Vol. 4854, Society of Photo-Optical Instrumentation Engineers (SPIE) Conference Series, ed. J.C.\,Blades, O.H.W.\,Siegmund, 129-140
\bibitem[\protect\citeauthoryear{Bressan et~al.}{2013}]{bressan2013}                         Bressan A., Marigo P., Girardi L., Nanni A., Rubele, S., 2013, 40th Li{\`e}ge International Astrophysical Colloquium ''Ageing Low Mass Stars: From Red Giants to White Dwarfs'', Li{\`e}ge, Belgium, Eds. J. Montalb{\'a}n, A. Noels, V. Van Grootel, EPJ Web of Conferences, Volume 43, id.03001
\bibitem[\protect\citeauthoryear{Brogaard et~al.}{2012}]{brogaard2012}                       Brogaard K. et~al., 2012, A\&A, 543, A106
\bibitem[\protect\citeauthoryear{Brogaard et~al.}{2014}]{brogaard2014}                       Brogaard K. et~al., 2014, 'Asteroseismology of Stellar Populations in the Milky Way', Ap\&SS Proc., Springer, Eds.: A.~Miglio, P.~Eggenberger, L.~Girardi, J~Montalban
\bibitem[\protect\citeauthoryear{Brown et~al.}{2011}]{brown2011}                             Brown T.~M., Latham D.~W., Everett M.~E., Esquerdo~G.~A., 2011, AJ, 142, 112 (http://cdsarc.u-strasbg.fr/viz-bin/Cat?V/133)
\bibitem[\protect\citeauthoryear{Bruntt et al.}{2010}]{bruntt2010}		 	     Bruntt H., et al., 2010, MNRAS, 405, 1907
\bibitem[\protect\citeauthoryear{Bruntt et al.}{2012}]{bruntt2012}			     Bruntt H., et al., 2012, MNRAS, 423, 122
\bibitem[\protect\citeauthoryear{Casagrande \& VandenBerg}{2014}]{casagrande2014}	     Casagrande L., VandenBerg D.~A., 2014, MNRAS 444, 392
\bibitem[\protect\citeauthoryear{Cesetti et~al.}{2013}]{cesetti2013}                         Cesetti M., Pizzella A., Ivanov V.~D., Morelli L., Corsini E.~M., Dalla Bont{\`a} E., 2013, A\&A, 549, A129
\bibitem[\protect\citeauthoryear{Corsaro et~al.}{2012}]{corsaro2012}                         Corsaro E. et~al., 2012, ApJ, 757, 190
\bibitem[\protect\citeauthoryear{Cutri et~al.}{2003}]{Cutri2003}                             Cutri R.~M. et~al., 2003, The 2MASS All-Sky Catalog of Point Sources, University of Massachusetts and Infrared Processing and Analysis Center, IPAC/California Institute of Technology
\bibitem[\protect\citeauthoryear{Debosscher et~al.}{2011}]{debosscher2011}                   Debosscher J., Blomme J., Aerts C., De Ridder J., 2011, A\&A 529, A89
\bibitem[\protect\citeauthoryear{Dias, Lepine \& Alessi}{2002}]{dias2002}                    Dias W.~S., Lepine J.~R.~D., Alessi B.~S., 2002, A\&A, 388, 168
\bibitem[\protect\citeauthoryear{D'Orazi et~al.}{2009}]{dorazi2009}                          D'Orazi V., Magrini L., Randich S., Galli D., Busso M., Sestito P., 2009, ApJ, 693, L31
\bibitem[\protect\citeauthoryear{Frandsen et~al.}{2013}]{frandsen2013}                       Frandsen S. et~al., 2013, A\&A, 556, A138
\bibitem[\protect\citeauthoryear{Frasca et~al.}{2003}]{Frasca2003}                           Frasca A., Alcal\`a J.~M., Covino E., Catalano S., Marilli, E., Paladino R., 2003, A\&A, 405, 149
\bibitem[\protect\citeauthoryear{Frasca et~al.}{2006}]{Frasca2006}                           Frasca A., Guillout P., Marilli E., Freire Ferrero R., Biazzo K., Klutsch A.,  2006, A\&A, 454, 301
\bibitem[\protect\citeauthoryear{Frinchaboy \& Majewski}{2008}]{frinchaboy2008}              Frinchaboy P.~M., Majewski S.~R., 2008, AJ, 136, 118
\bibitem[\protect\citeauthoryear{Glushkova, Batyrshinova \& Ibragimov}{1999}]{glushkova1999} Glushkova E.~V., Batyrshinova V.~M., Ibragimov M.~A., 1999, Astronomy Letters, 25, 86
\bibitem[\protect\citeauthoryear{G\'omez Maqueo Chew et~al.}{2013}]{gomez2013}               G\'omez Maqueo Chew Y. et~al., 2013, ApJ, 768, 79
\bibitem[\protect\citeauthoryear{Gonzalez et~al.}{2009}]{gonzalez2009}                       Gonzalez O.~A. et~al., 2009, A\&A, 508, 289
\bibitem[\protect\citeauthoryear{Grevesse \& Sauval}{1998}]{grevesse98}                      Grevesse N., Sauval A.~J., 1998, Sp.\, Sci.\, Rev., 85, 161
\bibitem[\protect\citeauthoryear{Grevesse, Asplund \& Sauval}{2007}]{grevesse07}             Grevesse N., Asplund M., Sauval A.~J., 2007, Sp.\,Sci.\,Rev., 130, 105
\bibitem[\protect\citeauthoryear{Grupp}{2004a}]{grupp2004a}                                  Grupp F., 2004, A\&A, 420, 289 
\bibitem[\protect\citeauthoryear{Grupp}{2004b}]{grupp2004b}                                  Grupp F., 2004, A\&A, 426, 309 
\bibitem[\protect\citeauthoryear{Gustafsson et~al.}{2008}]{gustafsson2008}                   Gustafsson B., Edvardsson B., Eriksson K., J{\o}rgensen U.~G., Nordlund {\AA}, Plez B., 2008, A\&A, 486, 951
\bibitem[\protect\citeauthoryear{Hekker et~al.}{2011}]{hekker2011}                           Hekker S. et~al., 2011, A\&A, 530, A100
\bibitem[\protect\citeauthoryear{Jacobson \& Friel}{2013}]{jacobson2013}		     Jacobson H.~R., Friel E.~D., 2013, AJ, 145, 107
\bibitem[\protect\citeauthoryear{Janes et~al.}{2013}]{janes2013}                             Janes K., Barnes S.\,A., Meibom S., Hoq S., 2013, AJ, 145,~7
\bibitem[\protect\citeauthoryear{Kharchenko et~al.}{2004}]{kharchenko2004}                   Kharchenko N.~V., Piskunov A.~E., R\"oser S., Schilbach E., Scholz R.-D., 2004, Astron. Nach., 325, 740
\bibitem[\protect\citeauthoryear{Kharchenko et~al.}{2005}]{kharchenko2005}                   Kharchenko N.~V., Piskunov A.~E., R\"oser S., Schilbach E., Scholz R.-D., 2005, A\&A, 438, 1163
\bibitem[\protect\citeauthoryear{Koch et~al.}{2010}]{koch2010}                               Koch D.~G. et~al.,, 2010, ApJ, 713, L79
\bibitem[\protect\citeauthoryear{Korn et~al.}{2003}]{korn2003}                               Korn A.~J., Shi J., Gehren T., 2003,  A\&A, 407, 691 
\bibitem[\protect\citeauthoryear{Kurucz}{1993}]{kurucz93}                                    Kurucz R., 1993, CD-ROM 18
\bibitem[\protect\citeauthoryear{Lind et~al.}{2012}]{lind2012}                               Lind K., Bergemann M., Asplund M., 2012, MNRAS, 427, 50
\bibitem[\protect\citeauthoryear{Lindoff}{1972}]{lindoff1972}                                Lindoff U., 1972, \aa, 16, 315
\bibitem[\protect\citeauthoryear{Luo et al.}{2009}]{luo2009}				     Luo Y.~P., Zhang X.~B., Luo C.~Q., Deng L.~C., Luo Z.~Q., 2009, New Astronomy 14, 584
\bibitem[\protect\citeauthoryear{Maiorca et~al.}{2011}]{maiorca2011}			     Maiorca E., Randich S., Busso M., Magrini L., Palmerini S., 2011, ApJ, 736, 120
\bibitem[\protect\citeauthoryear{Mermilliod \& Mayor}{1990}]{mermilliod1990}                 Mermilliod J.~C., Mayor M., 1990, A\&A, 237, 61
\bibitem[\protect\citeauthoryear{Mermilliod, Mayor \& Udry}{2008}]{mermilliod2008}           Mermilliod J.~C., Mayor M., Udry S., 2008, A\&A, 485, 303
\bibitem[\protect\citeauthoryear{Metcalfe et~al.}{2010}]{metcalfe2010}                       Metcalfe T.~S. et~al., 2010, ApJ, 723, 1583
\bibitem[\protect\citeauthoryear{Miglio et~al.}{2012}]{miglio2012}                           Miglio A. et~al., 2012, MNRAS, 419, 2077
\bibitem[\protect\citeauthoryear{Miglio et~al.}{2013}]{miglio2013}                           Miglio A. et~al., 2013, 40th Li{\`e}ge International Astrophysical Colloquium ''Ageing Low Mass Stars: From Red Giants to White Dwarfs'', Li{\`e}ge, Belgium, Eds. J. Montalb{\'a}n, A. Noels, V. Van Grootel, EPJ Web of Conferences, Volume 43, id.03004
\bibitem[\protect\citeauthoryear{Molenda-\.Zakowicz et~al.}{2010}]{molenda2010}              Molenda-\.Zakowicz J., Jerzykiewicz M., Frasca A., Catanzaro G., Kopacki G., Latham D.~W., 2010, arXiv~1005.0985
\bibitem[\protect\citeauthoryear{Molenda-\.Zakowicz et~al.}{2013}]{molenda2013}              Molenda-\.Zakowicz J. et~al., 2013, MNRAS, 434, 1422
\bibitem[\protect\citeauthoryear{Montalb{\'a}n et~al.}{2013}]{Montalban2013}                 Montalb{\'a}n J., Miglio A., Noels A., Dupret M.-A., Scuflaire R., Ventura P., 2013, ApJ, 766, 118
\bibitem[\protect\citeauthoryear{Niemczura, Morel \& Aerts}{2009}]{niemczura09}              Niemczura E., Morel T., Aerts C., 2009, A\&A, 506, 213
\bibitem[\protect\citeauthoryear{{\"O}nehag et~al.}{2011}]{onehag2011}                       {\"O}nehag A., Korn A., Gustafsson B., Stempels E., Vandenberg D.~A., 2011, A\&A, 528, A85
\bibitem[\protect\citeauthoryear{Pancino et~al.}{2010}]{pancino2010}                         Pancino E., Carrera R., Rossetti E., Gallart C., 2010, A\&A, 511, A56
\bibitem[\protect\citeauthoryear{Prugniel \& Soubiran}{2001}]{prugniel2001}                  Prugniel Ph., Soubiran C., 2001, A\&A, 369, 1048
\bibitem[\protect\citeauthoryear{Pr\u{s}a et~al.}{2011}]{prsa2011}                           Pr\u{s}a A. et~al., 2011, AJ, 141, 83
\bibitem[\protect\citeauthoryear{Salaris et~al.}{2004}]{salaris2004}			     Salaris M., Weiss A., Percival S.~M., 2004, A\&A, 414, 163
\bibitem[\protect\citeauthoryear{Sanders}{1971}]{sanders1971}                                Sanders W.~L., 1971, A\&A, 15, 368
\bibitem[\protect\citeauthoryear{Sandquist et~al.}{2013}]{sandquist2013}                     Sandquist E.~L. et~al., 2013, AJ, 146, 40
\bibitem[\protect\citeauthoryear{Santos, Israelian \& Mayor}{2004}]{santos2004}		     Santos N.~C., Israelian G., Mayor M., 2004, A\&A, 415, 1153
\bibitem[\protect\citeauthoryear{Sbordone}{2005}]{sbordone05}                                Sbordone L., 2005, MSAIS, 8, 61
\bibitem[\protect\citeauthoryear{Schlegel et~al.}{1998}]{schlegel1998}                       Schlegel D.~J., Finkbeiner D.~P., Davis M., 1998, ApJ, 500, 525
\bibitem[\protect\citeauthoryear{Shi et~al.}{2014}]{shi2014}                                 Shi J.~R., Gehren T., Zeng J.~L., Mashonkina L., Zhao G., 2014, ApJ, 782, 80 
\bibitem[\protect\citeauthoryear{Sousa et~al.}{2006}]{sousa2006} 			     Sousa S.~G., Santos N.~C., Israelian G., Mayor M., Monteiro M.~J.~P.~F.~G., 2006, A\&A, 458, 873
\bibitem[\protect\citeauthoryear{Sousa et~al.}{2008}]{sousa2008}			     Sousa S.~G., et~al., 2008, A\&A, 487, 373
\bibitem[\protect\citeauthoryear{Sousa et~al.}{2011a}]{sousa2011a}			     Sousa S.~G., et~al., 2011a, A\&A, 526, A99
\bibitem[\protect\citeauthoryear{Sousa et~al.}{2011b}]{sousa2011b}                           Sousa S.~G., et~al., 2011b, A\&A, 533, A141
\bibitem[\protect\citeauthoryear{Stello et~al.}{2009}]{stello2009}                           Stello D., Chaplin W.~J., Basu S., Elsworth Y., Bedding T.~R., 2009, MNRAS, 400, L80
\bibitem[\protect\citeauthoryear{Stello et~al.}{2011}]{stello2011}                           Stello D. et~al., 2011, ApJ, 739, 13
\bibitem[\protect\citeauthoryear{Stello et~al.}{2013}]{stello2013}                           Stello D. et~al., 2013, ApJ Letters, 765, L41
\bibitem[\protect\citeauthoryear{Thygesen et al.}{2012}]{thygesen2012}			     Thygesen A.~O., et al., 2012, \aa, 543, 160
\bibitem[\protect\citeauthoryear{Tonry \& Davis}{1979}]{tonry79}                             Tonry J., Davis M., 1979, AJ, 84, 1511
\bibitem[\protect\citeauthoryear{Topping}{1972}]{topping72}                                  Topping J., 1972, Errors of Observation and Their Treatment, Chapman Hall Ltd., 92
\bibitem[\protect\citeauthoryear{Udry et~al.}{1999}]{udry1999}                               Udry S. et~al., 1999, in Precise Stellar Radial Velocities ASP Conf.\ Ser., 185, 383
\bibitem[\protect\citeauthoryear{Uytterhoeven et~al.}{2010b}]{Uytterhoeven2010a}             Uytterhoeven K. et~al., 2010a, AN, 331, poster P30 (arXiv:1003.6089)
\bibitem[\protect\citeauthoryear{Uytterhoeven et~al.}{2010a}]{Uytterhoeven2010b}             Uytterhoeven K. et~al., 2010b, AN, 331, 993
\bibitem[\protect\citeauthoryear{Uytterhoeven et~al.}{2011}]{uytterhoeven2011}               Uytterhoeven K. et~al., 2011, A\&A, 534, 125
\bibitem[\protect\citeauthoryear{Valdes et~al.}{2004}]{valdes2004}                           Valdes F., Gupta R., Rose J.~A., Singh H.~P., Bell D.~J., 2004, ApJS, 152, 251
\bibitem[\protect\citeauthoryear{Valenti \& Piskunov}{1996}]{valenti1996}                    Valenti J.~A., Piskunov N., 1996, A\&AS, 118, 595
\bibitem[\protect\citeauthoryear{Van Cauteren et~al.}{2005}]{vancauteren2005}                Van Cauteren P., Lampens P., Robertson C.~W., Strigachev A., 2005. CoAst, 146, 21
\bibitem[\protect\citeauthoryear{Villanova et~al.}{2010}]{villanova2010}                     Villanova S., Randich S., Geisler D., Carraro G., Costa E., 2010, A\&A, 509, A\,102
\bibitem[\protect\citeauthoryear{Watson, Henden \& Price}{2006}]{Watson2006}                 Watson C., Henden A.~A., Price A., 2006, The AAVSO International Variable Star Index (Version 2011-04-17), Soc.\ for Astron.\ Sci.\ Ann.\ Symp.\ 25, 47
\bibitem[\protect\citeauthoryear{Wong \& Meibom}{2009}]{wong2009}                            Wong A., Meibom S., 2009, American Physical Society, Joint Fall 2009 Meeting of the Ohio Sections of the APS and AAPT, October 9-10, 2009, poster P1.006
\bibitem[\protect\citeauthoryear{Young et~al.}{2005}]{yong2005}				     Yong D., Carney B.~W., Teixera de Almeida M.~L., 2005, AJ, 130, 597

\end{thebibliography}
\end{document}